\documentclass[a4paper,11pt]{article}

\usepackage{balance}
\usepackage[font={small}, labelsep={period},labelfont={bf}]{caption}

\usepackage[utf8]{inputenc}
\usepackage[T1]{fontenc}
\usepackage{graphicx}
\usepackage{epsfig}
\usepackage[english]{babel}
\usepackage{listings}
\usepackage{hyperref}
\usepackage{authblk}
\usepackage{fancyhdr}
\usepackage{floatflt}
\usepackage{float}
\usepackage{wrapfig}
\usepackage{rotating}
\usepackage{longtable}
\usepackage{chngcntr}
\usepackage{relsize}
\usepackage{amsmath}
\usepackage{amsfonts}
\usepackage{amssymb}
\usepackage[amssymb, cdot, mediumqspace, thickqspace]{SIunits}
\usepackage[a4paper,top=3cm,bottom=3cm,left=2.5cm,right=2.5cm]{geometry}
\usepackage{mathrsfs}
\usepackage{booktabs}
\usepackage{braket}
\usepackage{subfigure}
\usepackage[titletoc, title]{appendix}
\usepackage{pdfpages}

\usepackage[style=phys, biblabel=brackets, citestyle=numeric-comp, backend=biber, sorting=none, doi=true, url=true, eprint=true, maxbibnames=2]{biblatex}

\usepackage[style=phys, biblabel=brackets, citestyle=numeric-comp, backend=biber, sorting=none, doi=true, url=true, eprint=true, maxbibnames=2]{biblatex}

\DeclareFieldFormat{titlecase}{#1} 
\DeclareFieldFormat{doi/url-link}{#1} 
\DeclareSourcemap{ 
  \maps[datatype=bibtex]{
    \map[overwrite]{
      \step[fieldsource=doi, final]
      \step[fieldset=eprint, null]
      \step[fieldset=url, null]
    }  
  }
}

\AtEveryBibitem{
    \clearfield{month}
}

\DeclareSourcemap{
 \maps[datatype=bibtex,overwrite=true]{
  \map{
   \pernottype{inproceeding}
    \step[fieldsource=Collaboration, final=true]
    \step[fieldset=usera, origfieldval, final=true]
  }
 }
}

\renewbibmacro*{note+pages}{
  \printfield{note}%
  \setunit{\bibpagespunct}%
  \printfield{pages}%
  \iffieldundef{pages}
    {%
      \clearfield{doi}%
    }%
    {%
    }%
}

\usepackage{placeins}
\usepackage{titlesec}
\usepackage{mathtools}
\usepackage{enumitem}
\usepackage{color}
\usepackage{mhchem}
\usepackage{slashed}
\usepackage{feynmp}
\usepackage{feynmp-auto}
\usepackage{lineno}
\usepackage[bottom]{footmisc}
\usepackage{tabularx}
\usepackage{lineno}
\usepackage{multirow}
\usepackage{tikz} 
\usetikzlibrary{shapes,arrows}
\usepackage{fancyvrb}
\usepackage{lscape}
\usepackage{comment}
\usepackage{cleveref}
\usepackage{orcidlink}
\usepackage{authblk}

\definecolor{denim}{rgb}{0.08, 0.38, 0.74}
\definecolor{darkblue}{rgb}{0.00, 0.00, 0.5}
\definecolor{internationalkleinblue}{rgb}{0.0, 0.18, 0.65}
\definecolor{bondiblue}{rgb}{0.0, 0.58, 0.71}
\definecolor{blue_ryb}{rgb}{0.01, 0.28, 1.0}
\definecolor{cobalt}{rgb}{0.0, 0.28, 0.67}
\definecolor{lightblue}{rgb}{0.145,0.6666,1}

\titleformat{\paragraph}
{\normalfont\normalsize\bfseries}{\theparagraph}{1em}{}
\titlespacing*{\paragraph}
{0pt}{3.25ex plus 1ex minus .2ex}{1.5ex plus .2ex}

\fancypagestyle{plain}{%
  \fancyhf{}\fancyfoot[C]{ \thepage}%
}

\hypersetup{
	colorlinks = true,
	allcolors = internationalkleinblue,
	linktocpage
}

\graphicspath{ {./figures/} }

\bibliography{bibliografia}

\begin{document}

\begin{figure}
\includegraphics[width=0.25\textwidth]{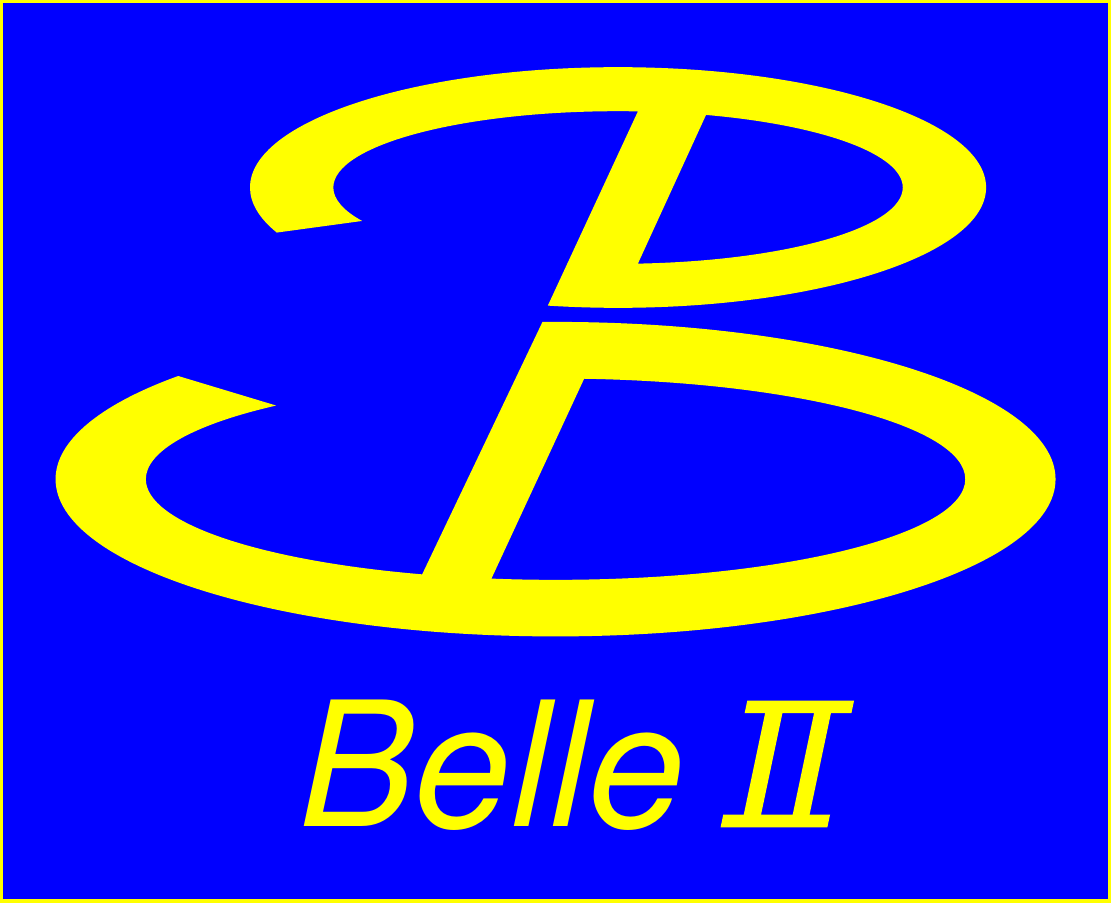}
\vspace{-3 cm}
\end{figure}

\begin{flushright}BELLE2-CONF-2023-003\\
May 2, 2023
\vspace{2 cm}
\end{flushright}

  \author{F.~Abudin{\'e}n\,\orcidlink{0000-0002-6737-3528}} 
  \author{I.~Adachi\,\orcidlink{0000-0003-2287-0173}} 
  \author{K.~Adamczyk\,\orcidlink{0000-0001-6208-0876}} 
  \author{L.~Aggarwal\,\orcidlink{0000-0002-0909-7537}} 
  \author{P.~Ahlburg\,\orcidlink{0000-0002-9832-7604}} 
  \author{H.~Ahmed\,\orcidlink{0000-0003-3976-7498}} 
  \author{J.~K.~Ahn\,\orcidlink{0000-0002-5795-2243}} 
  \author{H.~Aihara\,\orcidlink{0000-0002-1907-5964}} 
  \author{N.~Akopov\,\orcidlink{0000-0002-4425-2096}} 
  \author{A.~Aloisio\,\orcidlink{0000-0002-3883-6693}} 
% ...
\affil{(The Belle~II Collaboration)}
\date{}
\title {\Large\textbf{Observation of ${B\to D^{(*)} K^- K^{0}_S}$ decays \\using the 2019--2022 Belle~II data sample}}

{\let\newpage\relax\maketitle}
    
\begin{abstract}
We present a measurement of the branching fractions of four $B^{0,-}\to D^{(*)+,0} K^- K^{0}_S$ decay modes. 
The measurement is based on data from SuperKEKB electron-positron collisions at the $\Upsilon(4S)$  resonance collected with the Belle II detector and corresponding to an integrated luminosity of ${362~\text{fb}^{-1}}$. 
The event yields are extracted from fits to the distributions of the difference between expected and observed $B$ meson energy to separate signal and background, and are efficiency-corrected as a function of the invariant mass of the $K^-K_S^0$ system.  We find the branching fractions to be
\begin{align*}
    \mathcal{B}(B^-\to D^0K^-K_S^0)=&(1.89\pm 0.16\pm 0.10)\times 10^{-4},\\
    \mathcal{B}(\overline B{}^0\to D^+K^-K_S^0)=&(0.85\pm 0.11\pm 0.05)\times 10^{-4},\\
    \mathcal{B}(B^-\to D^{*0}K^-K_S^0)=&(1.57\pm 0.27\pm 0.12)\times 10^{-4},\\
    \mathcal{B}(\overline B{}^0\to D^{*+}K^-K_S^0)=&(0.96\pm 0.18\pm 0.06)\times 10^{-4},
\end{align*} 
where the first uncertainty is statistical and the second systematic. These results include the first observation of $\overline B{}^0\to D^+K^-K_S^0$, $B^-\to D^{*0}K^-K_S^0$, and $\overline B{}^0\to D^{*+}K^-K_S^0$ decays and a significant improvement in the precision of $\mathcal{B}(B^-\to D^0K^-K_S^0)$ compared to previous measurements.
\end{abstract}

\clearpage
\section{Introduction and motivation}

At the SuperKEKB energy-asymmetric $e^+e^-$ collider, bottom-antibottom meson pairs ($B\overline B$) are produced at threshold from $\Upsilon(4S)$ decays. A large part of the Belle~II physics program~\cite{Belle2:PhysicsBook} relies on identifying the partner $B$ meson produced in association with the signal $B$ meson, to infer the properties of the signal ($B$-tagging). 
In particular, in hadronic $B$-tagging the partner $B$ meson is fully reconstructed to infer the kinematic properties of the signal using initial-state constraints. 
The Belle~II $B$-tagging algorithm, full event interpretation (FEI), is based on a set of multivariate classifiers trained on the Belle~II Monte Carlo (MC) simulation~\cite{Belle2:FEI}. A mismodeling in the simulation may introduce biases and larger uncertainties in the FEI efficiencies, and leads to a suboptimal FEI performance, degrading the Belle~II reach. Knowledge of $B$ meson hadronic decays is limited: about $40\%$ of the total $B$ width is not measured in terms of exclusive branching fractions ($\mathcal{B}$), and thus is generated by the simulation with the \texttt{Pythia} fragmentation model~\cite{MC:pythia}, which is known to be inaccurate.  The total branching fraction of $B\to D^{(*)}K^{(*)}K^{(*)}$ could be as large as 6\%, according to \texttt{Pythia} (where $D$ and $K$ stands for both neutral and charged mesons). However, only a small fraction of the exclusive components has been measured~\cite{Belle:DKK} and currently these are not used by the FEI algorithm. The high purity of these $B\to D^{(*)}K^{(*)}K^{(*)}$ decays makes them ideal candidates to improve the $B$-tagging efficiency. Therefore, an improvement in the description of these $B$ branching fractions can lead to an improvement of the FEI efficiency. Moreover, the branching fractions are not sufficient, and knowledge of the  final-state kinematic properties and the intermediate states is essential for an accurate description of these decays when including them into the Belle~II $B$-tagging algorithm.

The ${B\to D^{(*)} K^- K^{0}_S}$ decays proceed via a tree-level amplitude with an external $W$ boson emission and the production of a $s\overline s$ pair. However, a quasi-two-body mechanism, $B\to D^{(*)}X^-(\to K^- K^{0}_S)$, with an intermediate resonance $X^-$, which decays strongly, is possible. In this case, a spin-parity assignment $J^P=1^-$ is expected for the $K^- K^{0}_S$ system if exact isospin symmetry is assumed~\cite{Diehl:spinparity}. The $X^-$ resonance is therefore a $\rho$-like particle, but the $\rho(770)$ is not favored given the limited available phase-space. Unfortunately, the resonant and non-resonant components cannot be easily identified and separated given the large width of the relevant resonances.

A search for ${B^{0,-}\to D^{(*)+,0} K^- K^{0(*)}}$ decays has been performed by the Belle experiment,  on a sample with an integrated luminosity of $29~\text{fb}^{-1}$~\cite{Belle:DKK}.
Four $B\to D^{(*)}K^-K^{*0}$ modes and the $B^-\to D^0 K^- K^0_S$ mode were observed, while excesses in the remaining $K_S^0$ modes were seen with significances of about 2.5~standard deviations. 
A study of the $m(K^-K^{*})$ invariant mass distribution was also reported by Belle in Ref.~\cite{Belle:DKK}. A dominant decay via the $a_1^-$ resonance was claimed for the ${B\to D^{(*)} K^- K^{*}}$ decays.
Due to the small sample size, no conclusive claims were made for the ${B\to D^{(*)} K^- K^{0}_S}$ decays. 
However, the $m(K^-K^{0}_S)$ invariant mass and angular distributions also suggested the presence of a resonant component in ${B\to D^{(*)} K^- K^{0}_S}$ decays.

A more precise measurement of the $B^-\to D^0 K^- K^0_S$ mode and the observation of the three remaining ${B\to D^{(*)} K^- K^{0}_S}$ modes is reported here, using the Belle~II data sample collected before the 2022 long shutdown (LS1), corresponding to an integrated luminosity of 362~fb$^{-1}$ on the $\Upsilon(4S)$ resonance. 

This analysis aims to better understand the intermediate states of the decays by investigating the $m(K^-K^{0}_S)$ distribution. 
The description of the resonances in these decays is a topic of interest within the theory community~\cite{theory:molecular_a1,theory:exotic_a1,theory:rho}. 

The branching fractions are extracted independently for each channel. 
After the event reconstruction and selection, a fit to the $\Delta E = E_B^*-E_\text{beam}^*$ distribution is performed to separate the signal from the backgrounds. The $*$ indicates that the variable is evaluated in the $\Upsilon(4S)$ center-of-mass frame; $E_B^*$ is the reconstructed $B$ meson energy and $E_\text{beam}^*$ is the calibrated value of the beam energy. 
The $s$Plot procedure~\cite{splot} is used to obtain a background-free $m(K^-K^{0}_S)$ distribution. The signal efficiency is evaluated using a simulated signal sample, as a function of $m(K^-K^{0}_S)$. 
The branching fraction is then obtained from the efficiency-corrected integral of the $m(K^-K^{0}_S)$ distribution. 
The ${B\to D^{(*)}D_s^-(\to K^-K^{0}_S)}$ and ${B^-\to D^{*0}\pi^-}$ control channels are used to validate the analysis and assess some of the systematic uncertainties. We blind the signal region and use the simulation and the control samples to develop and optimize the procedures and selections.   
We use natural units $\hbar=c=1$ and charge-conjugation is implied throughout.

\section{The Belle~II detector and samples}\label{sec:Belle2}

The Belle~II detector~\cite{Belle2:TDR} is located at the SuperKEKB~\cite{SuperKEKB:TDR} energy-asymmetric electron-positron collider. 
It consists of several subdetectors arranged around the interaction point (IP) in a cylindrical geometry. 
The longitudinal direction ($z$ axis) is approximately aligned with the beam direction, while the $x$ and $y$ axes lie on the transverse plane, pointing horizontally outside the ring and vertically upward, respectively, with the radial displacement $r=\sqrt{x^2+y^2}$. The polar angle $\theta$ and the azimuthal angle $\phi$ are defined with respect to the $z$ axis.

The central feature of Belle~II is a superconducting solenoid, providing an inner uniform $1.5~\text{T}$ magnetic field, aligned to the longitudinal direction. 
The tracking system consists of two layers of pixel detector, four layers of double-sided strip detector, and a 56-layer  drift chamber~(CDC). The outer pixel layer is incomplete and only covers 15\% of the azimuthal surface. The CDC also has a central role in the trigger system.  
A time-of-propagation detector and an aerogel ring-imaging Cherenkov detector are placed outside the tracking system in the central $\theta$ region (barrel, $31^\circ<\theta<128^\circ$) and forward $\theta$ region (endcap,  $14^\circ<\theta<30^\circ$), respectively. Both particle-identification detectors exploit Cherenkov light emission. 

The ECL is the outermost subdetector within the solenoid, a homogeneous electromagnetic calorimeter composed of \text{CsI(Tl)} crystals. The main tasks of the ECL are the measurement of the energies of photons, electrons, and neutral kaons and providing inputs for particle identification and to the trigger system. 
Outside the solenoid is the $K_L$ and $\mu$ detector, which consists of resistive-plate chambers and  scintillator strips. The active layers are interleaved with iron plates that form the return yoke of the magnetic field.

Belle~II integrates a hardware and software trigger to cope with the beam crossing rate of SuperKEKB, to suppress the background rate and to select with high efficiency $b\overline b$, $c\overline c$, and $\tau^+\tau^-$ events. 

The full data sample of Belle~II collected before LS1 is used in this analysis. The total integrated luminosity of the sample is $\mathcal L_\text{int}=362\pm 2\,\text{fb}^{-1}$, which corresponds to $(387\pm 6)\times 10^6$ $B\overline B$ pairs.

The data are processed using the Belle II analysis software~\cite{Belle2:basf2}. Simulated
events are generated using \texttt{KKMC} for quark-antiquark pairs from $e^+e^-$ collisions~\cite{MC:KKMC}, \texttt{Pythia8}
for hadronization~\cite{MC:pythia}, \texttt{EvtGen} for the decay of the generated hadrons~\cite{MC:evtgen}, and \texttt{Geant4} for the
detector response~\cite{MC:geant4}. The simulation includes simulated beam-induced backgrounds~\cite{Belle2:bkg}. 

Simulated Monte Carlo samples of the specific signals studied in this analysis are used to determine signal efficiencies, to define fit models, and to evaluate the systematic uncertainties. Each of the three-body decay modes ${B^{0,-}\to D^{(*)+,0} K^- K_S^0}$ is generated according to a phase-space distribution. A simulated sample that reproduces the composition of Belle~II events, including $B\overline B$ and $e^+e^-\to q\overline q$ (where $q$ indicates an $u,d,s$ or $c$ quark) backgrounds, and equivalent to an integrated luminosity of $1~\text{ab}^{-1}$, is used to investigate the sample composition and validate the analysis before unblinding of the signal region. 

\section{Reconstruction and event selection}\label{sec:reco}

Candidate charged pions and kaons are identified from their trajectories (tracks), which are required to have transverse and longitudinal impact parameters ${d_r<2~\text{cm}}$ and ${|d_z|<4~\text{cm}}$, respectively, with respect to the IP, $\theta$ angle within the acceptance of the CDC, and at least 20 hits in the CDC. They are identified as pions or kaons based on the ratio between the particle-identification (PID) likelihood for the test hypothesis and the sum of the likelihoods for all other hypotheses. The PID selection efficiency ranges between 50\% and 80\%, with misidentification probability ranging between 10\% and 3\%,  depending on the $p_T$ and $\cos\theta$ of the charged particle, both for kaon- and pion-enriched samples. 

Photons are selected by requiring ECL energy-deposits (clusters) involving more than one crystal with polar angle in the CDC acceptance, and a polar-angle-dependent energy threshold of 80~MeV, 30~MeV, and 60~MeV in the forward endcap, barrel, and backward endcap, respectively. 
Clusters are required to be detected within 200~ns from the beam crossing to suppress photons from beam background. The misreconstructed photons generated by hadronic clusters without a matched track are suppressed with a dedicated multivariate classifier, which uses cluster-shape variables, crystal pulse-shape, and track-cluster distance.

Neutral pions are reconstructed from photon pairs, requiring the diphoton mass to be between $120~\text{MeV}$ and $145~\text{MeV}$ and the convergence of a mass-constrained fit.  
Neutral kaons are reconstructed from pairs of opposite-charge pions. We require the dipion mass to be within  $10~\text{MeV}$ of the nominal value, a successful vertex fit,  the transverse distance between the IP and the pions vertex to exceed 0.4~cm, and the cosine of the angle between the $K_S^0$ flight direction and its reconstructed momentum to exceed 0.8.

Candidate $D^{(*)}$ mesons are reconstructed using the $D^0\to K^-\pi^+$, $D^+\to K^-\pi^+\pi^+$, $D^{*0}\to D^0\pi^0$, and $D^{*+}\to D^0\pi^+$ decay channels. The invariant mass of the $D^0$ and $D^+$ candidates is required to be within $15~\text{MeV}$ of the nominal value (corresponding to $3\sigma$, where $\sigma$ is the mass resolution) and a mass- and vertex-constrained fit is performed. Similarly, the invariant mass of the $D^{*0}$ ($D^{*+}$) is required to be within $3~\text{MeV}$ ($1.5~\text{MeV}$) of the nominal value  ($3\sigma$).

Candidates $B$ are reconstructed in the four modes  ${B^{0,-}\to D^{(*)+,0} K^- K^{0}_S}$. 
In all the modes we require ${M_{\rm bc}=\sqrt{E^{*2}_\text{beam}-p^{*2}_B}>5.272~\text{GeV}}$ and ${-0.12~\text{GeV}<\Delta E<0.3~\text{GeV}}$, where $p_B$ is the momentum of the $B$ meson.
The latter asymmetric requirement suppresses feed-across from other channels in which a pion is missing in the reconstruction. 
To suppress the continuum background, we require that the ratio of the second to the zeroth order Fox-Wolfram moment satisfies $R_2<0.5$~\cite{theory:FWM}; the absolute value of the cosine of the angle between the thrust axis of the $B$ and the thrust axis of the rest-of-event is smaller than 0.85, where the thrust axis is the direction that maximizes the total projection of the momenta of all particles in the event; and the absolute value of the cosine of the angle between the $B$ momentum and the beam direction satisfies ${|\cos\theta_{p_B^*p_\text{beam}^*}|<0.9}$. 

The rest-of-event includes all the remaining tracks and clusters selected with loose requirements. The tracks are required to be within the CDC polar acceptance, to have transverse momenta greater than $100~\text{MeV}$, and to have $d_r<0.5~\text{cm}$, $|d_z|<3~\text{cm}$; the clusters are required to be in the CDC polar acceptance, to have energies above $50~\text{MeV}$, and to be detected within 200~ns from the beam crossing. 

We also require $m(K^-K^{0}_S)$ to differ by more than 20~MeV  ($4\sigma$) from the nominal $D_s^-$ mass, to suppress resonant $B\to DD_s(\to K^-K^{0}_S)$ background, which has the same final state as the signal events. 

The average $B$ candidate multiplicity per event is between $1.05$ and $1.08$ for the $D^0$, $D^+$, and $D^{*+}$ channels,  while it is about $1.4$ for the $D^{*0}$ channel. 
For each event, the $B$ candidate with $M_{\rm bc}$ closest to the nominal $B$ mass is selected. 
The efficiency of this candidate selection, defined as the ratio of the number of correct candidates chosen to the total number of candidates for the events in which the correct candidate is reconstructed, is  above 92\% for all channels, based on  simulation. 
After the candidate selection, the purity of the sample (defined as the ratio of correctly reconstructed events to the total number of events) is above 90\% for all channels, except for the $D^{*0}$ channels where it is 71\%, based on the simulation.

We apply corrections for the track-momentum scale, photon-energy scale, and beam-energy scale, which are derived from large data control samples. 

\section{Yield extraction}\label{sec:yield}

The composition of the selected samples is investigated using the MC simulation. A small smooth $e^+e^-\to q\overline q$ continuum background is expected in all channels. In the ${\overline B{}^0\to D^+K^- K_S^0}$ channel a feed-across component from the ${\overline B{}^0\to D^{*+}K^-  K_S^0}$ channel is expected, in which the $\pi^0$ from the $D^{*+}$ decay is not reconstructed. This component peaks at $\Delta E=-0.15~\text{GeV}$. The same effect is present in the $B^-\to D^0K^-K_S^0$ channel with a feed-across component from the $B^-\to D^{*0}K^-K_S^0$ and $\overline B{}^0\to D^{*+}K^-  K_S^0$ channels, which both mimic the signal when the $\pi^0$ or the $\pi^+$ of the $D^*$ is not reconstructed. 
In the $B^-\to D^{*0}K^-K_S^0$ channel, a feed-across component from the $B^-\to D^0K^-K_S^0$ channel is reconstructed if an incorrectly reconstructed $\pi^0$ is associated to the $D^0$. This background component peaks around $\Delta E=0.15~\text{GeV}$. In the $B^-\to D^{*0}K^-K_S^0$ channel a feed-across component from the $\overline B{}^0\to D^{*+} K^- K^{0}_S$ channel is also expected, when the $\pi^0$ from the $D^{*0}$ decay is not reconstructed and an incorrectly reconstructed $\pi^+$ is associated. This background component peaks under the signal and requires a special treatment. The $\overline B{}^0\to D^{*+} K^- K^{0}_S$ channel has no peaking backgrounds. 

The \texttt{RooFit} package~\cite{ROOT:roofit} is used to perform extended maximum likelihood fits to the unbinned $\Delta E $ distribution.  The fit is performed in the $\Delta E$ range  ${[-0.12~\text{GeV}, 0.3~\text{GeV}]}$ for all channels. 
The range is limited on the lower side to exclude the missing-pion feed-across components.

The range extends up to $\Delta E=0.3~\text{GeV}$ to increase the robustness of the fit for the $B^- \to D^0 K^- K^0_S$, $\overline B{}^0 \to D^{+} K^- K^{0}_S$,  and $\overline B{}^0 \to D^{*+} K^- K^0_S$ channels. 
In the $B^- \to D^{*0} K^- K^0_S$ channel the extended range is used to constrain the peaking background.

In the $B^- \to D^0 K^- K^0_S$, $\overline B{}^0 \to D^{+} K^- K^{0}_S$,  and $\overline B{}^0 \to D^{*+} K^- K^0_S$ channels, the signal is described with the sum of one symmetric (\textit{core}) and one asymmetric (\textit{tail}) Gaussian distribution. The means of the two Gaussian distributions are constrained to be equal. The widths are fixed to the values obtained in a fit to the simulated signal sample.  However, a "fudge factor" $F$ is assigned as a multiplier to the core Gaussian width. The fudge factor is a floating parameter in the $D^{0}$  channel fit. In the $D^{+}$ and $D^{*+}$ channels, the fudge factors are fixed to the value from the $D^0$ fit ($F=1.1\pm 0.1$), since the expected yields are too small to allow for an additional floating parameter. The fractions of the core and the tail Gaussian are fixed to the values obtained in a fit to the signal simulation sample.  The background is described by the sum of a falling exponential distribution and a constant. The fit has two or three floating shape parameters (mean, exponential constant, and fudge factor for $D^0$ channel), plus the signal yield, and two background yields. 

The ${B^- \to D^{*0} K^- K^0_S}$ channel parameterization requires a different approach, because of the peaking background from the ${\overline B{}^0 \to D^{*+} K^- K^0_S}$ channel. 
The yield of this feed-across component can be constrained directly from data using the feed-across from the $B^- \to D^{0} K^- K^0_S$ channel. 
The latter component is shifted in $\Delta E$ by ${0.15~\text{GeV}}$. 
This $\Delta E$ region is almost continuum-background-free. The yield $N_{D^0}^\text{bkg}$ of feed-across from the $B^- \to D^{0} K^- K^0_S$ channel can be determined from data as 
\begin{equation}
N_{D^0}^\text{bkg}\propto \mathcal{B}(B^- \to D^{0} K^- K^0_S)f_{\pi^0}\varepsilon_{ B^- \to D^{0} K^- K^0_S},
\end{equation} 
where $\varepsilon$ is the efficiency and $f_{\pi^0}$ is the probability of incorrect $\pi^0$ association. For the $\overline B{}^0 \to D^{*+} K^- K^0_S$ feed-across we have 
\begin{equation}
N_{D^{*+}}^\text{bkg}\propto \mathcal{B}(\overline B{}^0 \to D^{*+} K^- K^0_S) f_{\pi^0}\varepsilon_{ D^{0} K^- K^0_S\text{ from }\overline B{}^0 \to D^{*+} K^- K^0_S}.
\end{equation}
Assuming
\begin{equation}\label{eq:Dst0_hypothesis}
\varepsilon_{D^{0} K^- K^0_S\text{ from }\overline B{}^0 \to D^{*+} K^- K^0_S}=\varepsilon_{ B^- \to D^{0} K^- K^0_S},
\end{equation}
we have
\begin{equation}\label{eq:DstpBkg_yield}
   N_{D^{*+}}^\text{bkg} = \frac{\mathcal{B}(\overline B{}^0 \to D^{*+} K^- K^0_S)}{\mathcal{B}(B^- \to D^{0} K^- K^0_S)} \times  N_{D^0}^\text{bkg}.
\end{equation}
Thus, by measuring the yield of the $\Delta E$-shifted feed-across from the $B^- \to D^{0} K^- K^0_S$ channel, and adding as inputs the branching fractions of the two feed-across backgrounds measured in this analysis, we constrain the yield of the peaking feed-across from $\overline B{}^0 \to D^{*+} K^- K^0_S$ decays. Equation~\ref{eq:Dst0_hypothesis} assumes that
 the efficiency of a multiparticle final state factorizes into the products of single-particle efficiencies. This assumption is justified since the decay chain of $\overline B{}^0 \to D^{*+} K^- K^0_S$ is identical to $\overline B{}^- \to D^{0} K^- K^0_S$ (except for the additional low-momentum $\pi^+$) and no major kinematic differences are expected in the $D^{0} K^- K^0_S$ channel that could lead to a difference in the efficiency. 
A systematic uncertainty is assigned due to this assumption.

The ${B^- \to D^{*0} K^- K^0_S}$ channel is fitted using the strategy described above. The functional forms previously described are used for signal and background distributions. 
 The fudge factor is fixed to unity, as estimated from the $B^-\to D^{*0}\pi^-$ control channel. 
 The feed-across component from the $B^- \to D^{0} K^- K^0_S$ channel is parameterized as an asymmetric Gaussian distribution with widths fixed from the signal simulation and mean equal to the signal mean with a fixed shift evaluated from the signal simulation. The $B^- \to D^{0} K^- K^0_S$ feed-across yield is free in the fit. 
 The feed-across component from the $\overline B{}^0 \to D^{*+} K^- K^0_S$ channel is fitted as an asymmetric Gaussian distribution with its mean fixed to zero and width fixed from the signal simulation. The yield is fixed based on the relation of Eq.~\ref{eq:DstpBkg_yield}, where $N_{D^0}^\text{bkg}$ is the yield of the $B^- \to D^{0} K^- K^0_S$ feed-across, which is fitted simultaneously, and the branching fractions fixed. The fits have two shape parameters, the signal yield, the yields of two backgrounds, and the yields of the $B^- \to D^{0} K^- K^0_S$ feed-across component as floating parameters.

\begin{figure}[!t]
\centering
\subfigure{\includegraphics[width=0.49\columnwidth]{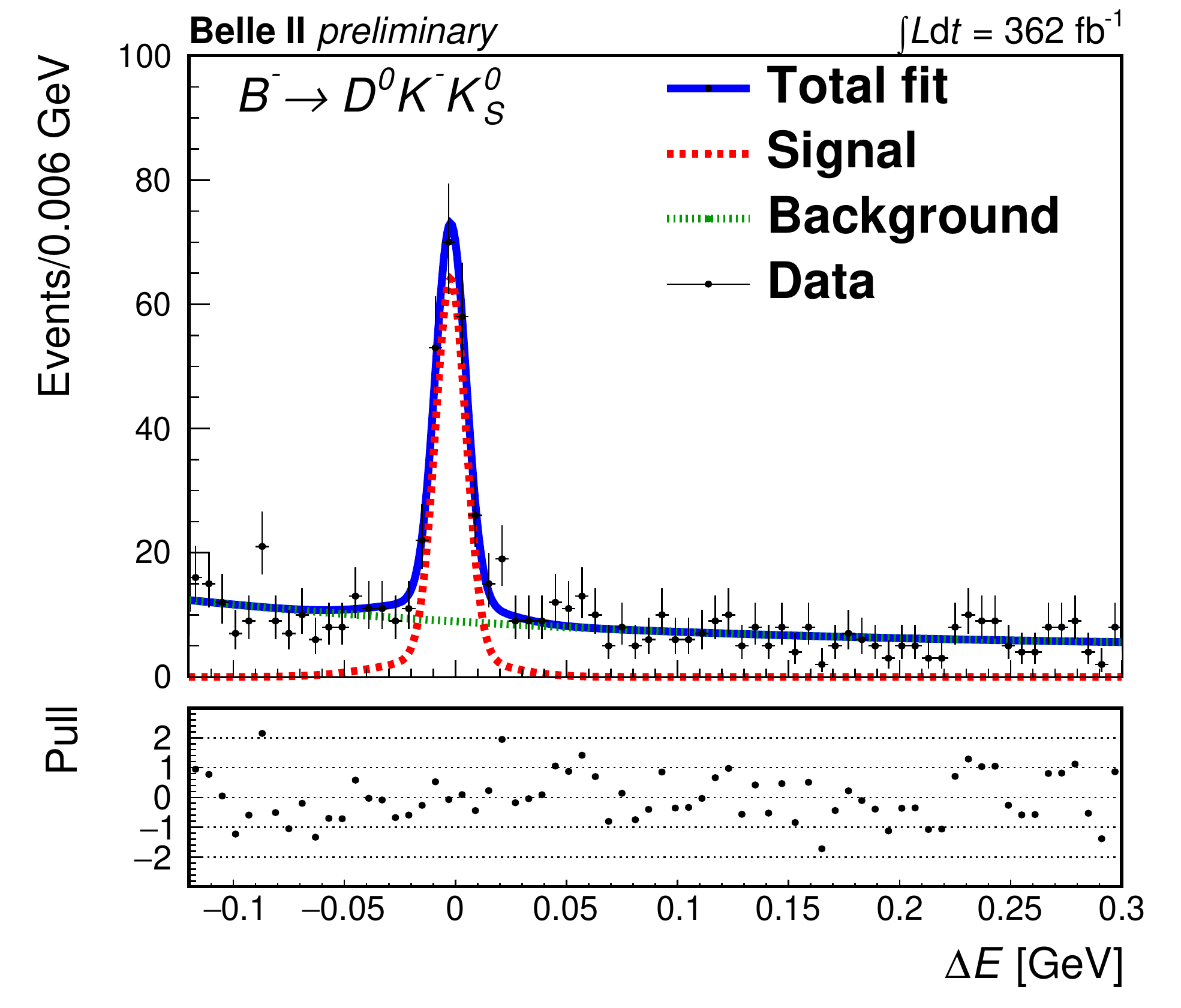}}
\subfigure{\includegraphics[width=0.49\columnwidth]{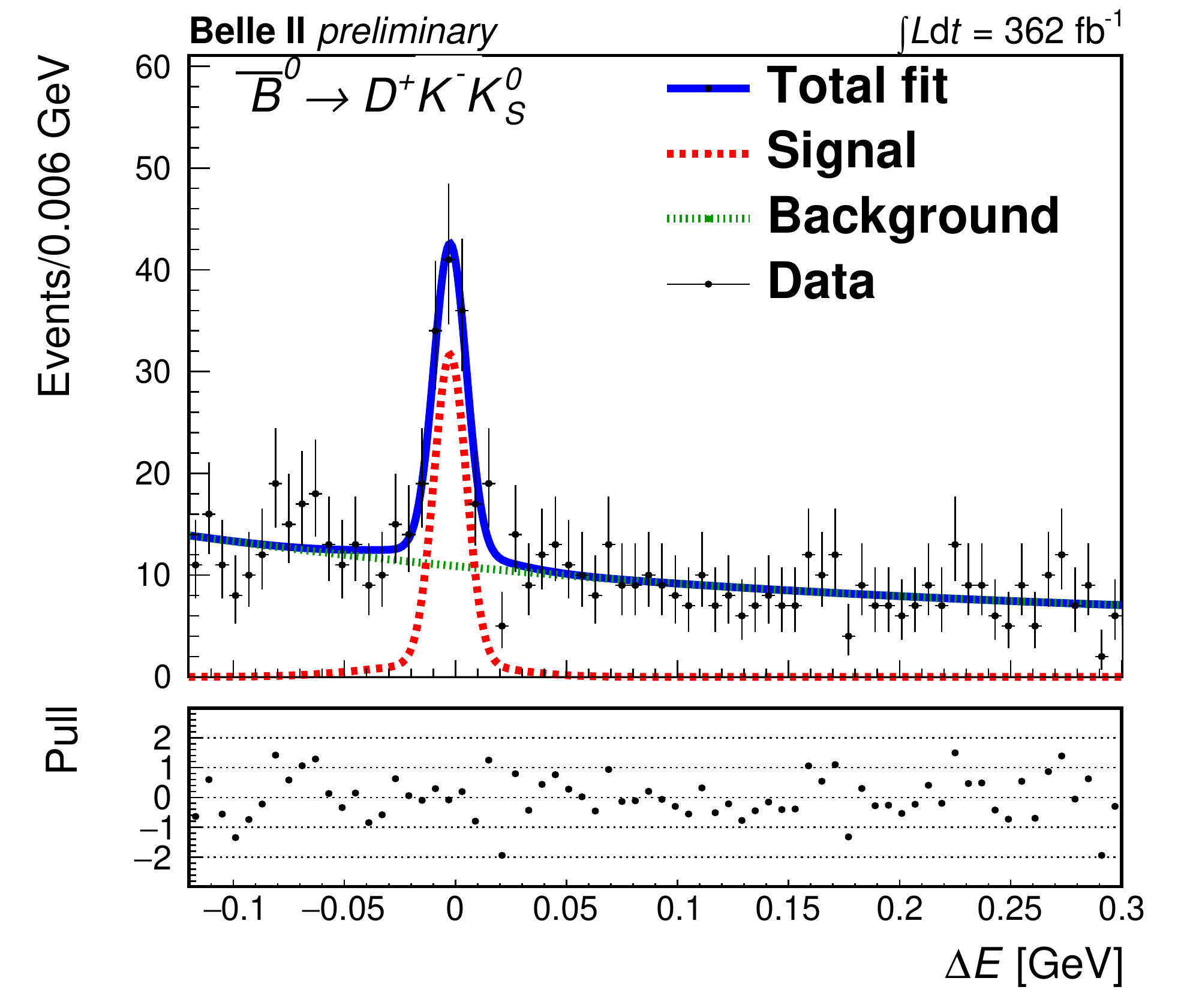}}
\subfigure{\includegraphics[width=0.49\columnwidth]{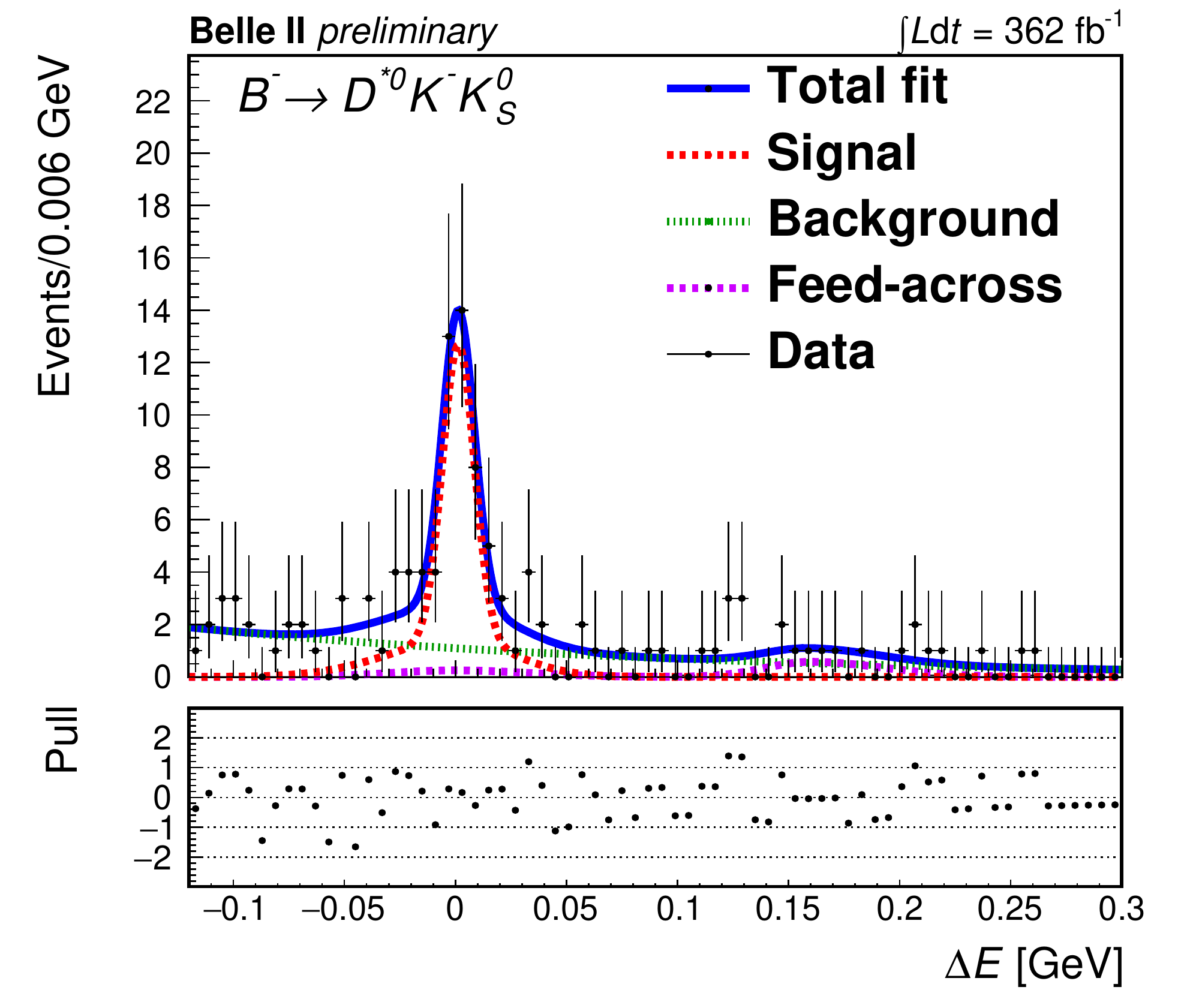}}
\subfigure{\includegraphics[width=0.49\columnwidth]{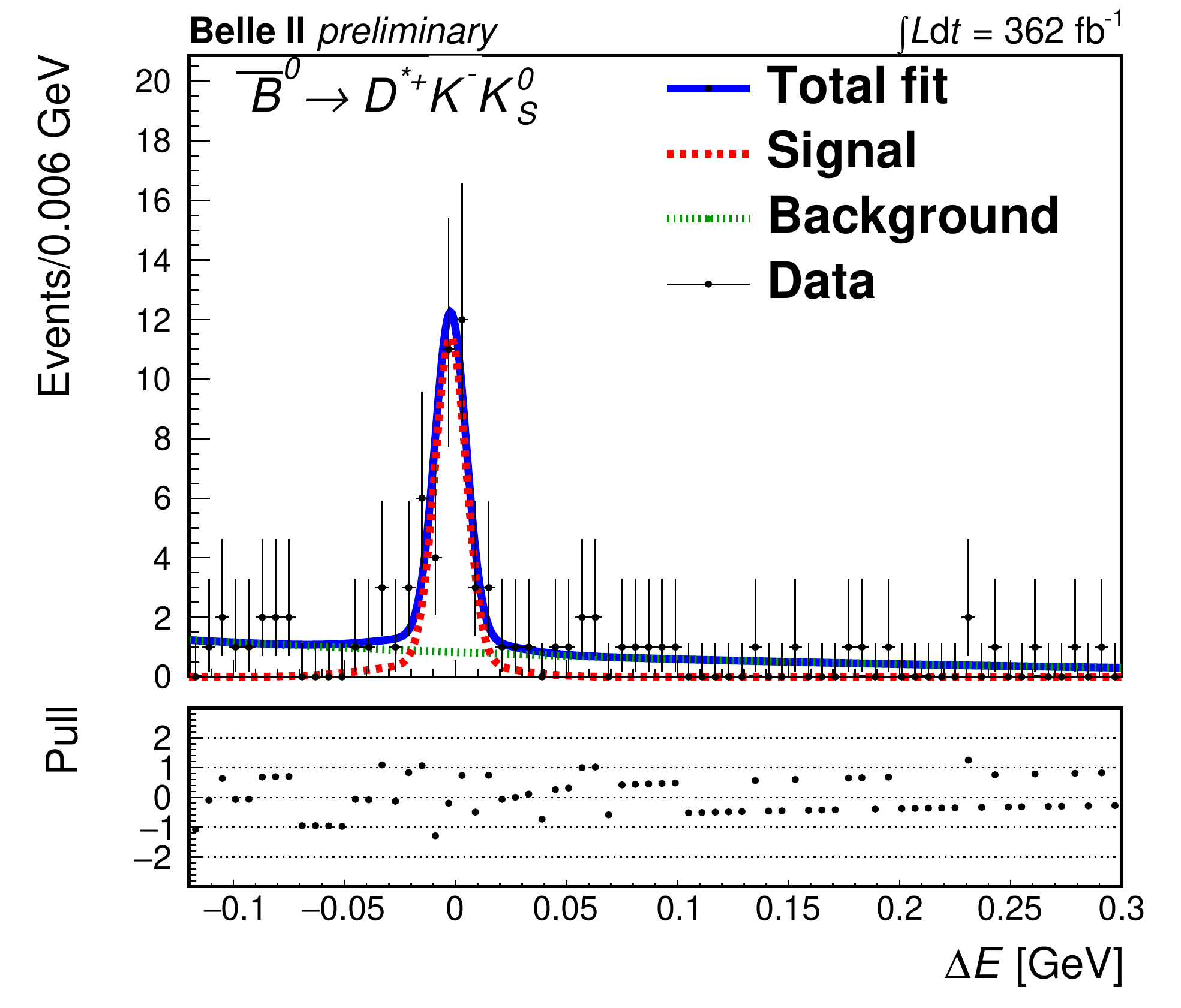}}
\caption{Distribution of $\Delta E$ for $B^-\to D^0K^-K_S^0$ (top left), $\overline B{}^0\to D^+K^-K_S^0$ (top right), $B^-\to D^{*0}K^-K_S^0$ (bottom left), and $\overline B{}^0\to D^{*+}K^-K_S^0$ (bottom right) channels, with the projection of fits overlaid. The fit components are highlighted, and the pulls between the fit and the data are shown below each distribution. } \label{fig:deltaE_fit_KS}
\end{figure}

 Data with fit projections overlaid are shown in Fig.~\ref{fig:deltaE_fit_KS}. Prominent signals of 207, 108, 51, and 37 events are observed for $D^0$, $D^+$, $D^{*0}$ and $D^{*+}$ channels, respectively. The backgrounds are smooth and small as expected, with a signal-to-background at peak between 4 and 14. The feed-across components in the $D^{*0}$ channel are well modeled. The four signals have statistical significances well above five standard deviations. 
 The pull between the data distribution and the fit is also shown, defined as the difference between fit value and the data yield divided by the data uncertainty. The agreement is acceptable.   The yields are summarized in Table~\ref{tab:BR_data}. 

The fits are validated by repeating the analysis on $10^3$ simulated pseudo-experiments, and they do not show any bias. 
The fit is validated on data using the ${B^{0,-}\to D^{(*)+,0} D_s^-(\to K^- K^{0}_S)}$ control channel, which shares the same final state as the signal, and a very similar decay chain. Thus, the efficiency is similar to the $B\to D^{(*)}K^-K_S^0$ one. In addition, the control modes are well measured~\cite{PDG}. To extract the control-channel signal, the selection procedure of Sec.~\ref{sec:reco} is used, however that the $m_{D_s^-}$ veto is inverted. The resulting control-channel branching fractions are in agreement with the expected values, showing that the fit procedure is unbiased. 

\section{Efficiency estimation}\label{sec:efficiency}

The efficiency is estimated for each channel separately using the signal simulation samples. 
The efficiency $\varepsilon(m_{KK})$ is defined as the fraction of the generated events that are reconstructed and selected in each bin of reconstructed $m(K^-K_S^0)$. 
This allows the efficiency to be independent of the $m(K^-K^{0}_S)$ distribution of the data, which is unknown and possibly different from the simulation. 
We divide the reconstructed $m(K^-K^{0}_S)$ distribution into 20 equally-spaced intervals (bins) between 1~GeV and 3.5~GeV. For each bin we determine the yield from a fit to the $\Delta E$ distribution, using the same functional forms described in Sec.~\ref{sec:yield}.

The efficiency is corrected for known data-simulation mismodelings. In particular, the PID selection efficiency is calibrated with a scale factor as a function of momentum and polar angle for each track; the $K_S^0$ reconstruction and selection efficiency is calibrated with a scale factor as a function of the vertex distance; the reconstruction and selection efficiencies of the low-momentum pions from the $D^{*+}$ and $D^{*0}$ decays are calibrated with two dedicated scale factors as a function of the pion momentum. The overall correction factor ranges between 1\% and 10\%. 

\begin{figure}[!hbt]
\centering
\includegraphics[width=0.6\columnwidth]{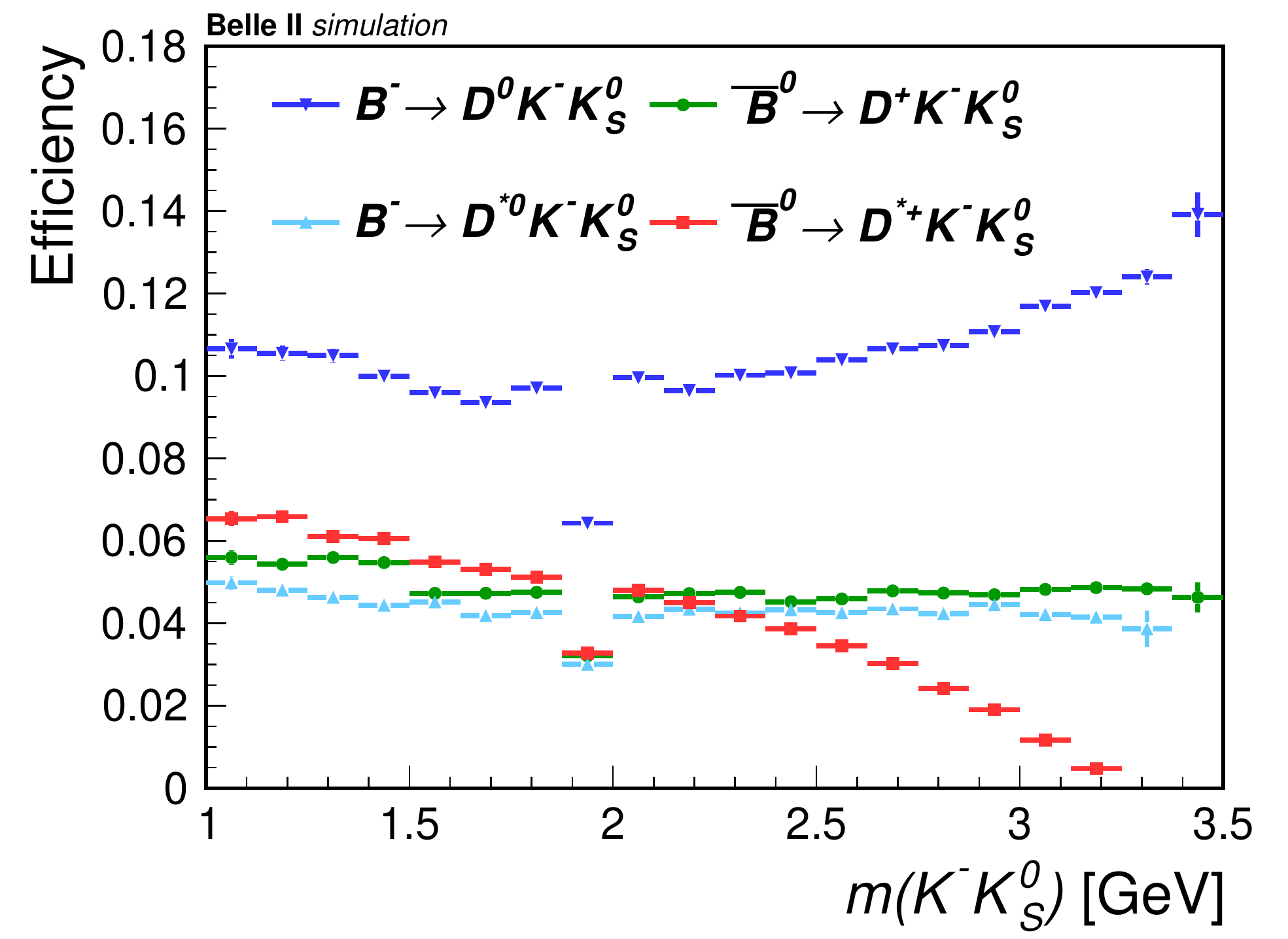}
\caption{Efficiency as a function of $m(K^-K^{0}_S)$ for the four decay channels evaluated on the signal simulation samples.}\label{fig:efficiency}
\end{figure}

The efficiencies for the four channels as functions of $m(K^-K^{0}_S)$ are shown in Fig.~\ref{fig:efficiency}. All channels show a drop in efficiency at $m(K^-K^{0}_S)\approx m_{D_s}$ due to the ${B\to DD_s^-}$ veto. 
The efficiency is stable down to $m(K^-K^{0}_S)=1~\text{GeV}$, which is the $K^-K_S^0$ invariant mass threshold. 
For the $D^{*+}$ channel, $\varepsilon(m_{KK})$ decreases at high $m(K^-K^{0}_S)$. This is due to the anti-correlation between $m(K^-K^{0}_S)$ and the $D^{*+}$-pion momentum, for which the efficiency decreases at low momentum. 

\section{Branching fraction extraction}\label{sec:BR}

The branching fraction can be expressed, including the dependence on $m(K^-K_S^0)$ of the yield and the efficiency, as

\begin{equation}\label{eq:BR}
    \mathcal{B}=\frac{1}{2 f_{+-,00} N_{B\overline B}  \mathcal{B}_{D^{(*)}}\mathcal{B}_{K_S^0}} \sum_{i=1}^{20}\frac{N_i^\text{reco}}{\varepsilon_i},
\end{equation}
where the index $i$ runs over $m(K^-K^{0}_S)$ bins, $N_i^\text{reco}$ and $\varepsilon_i$ are the yield and the efficiency in each bin, $\mathcal{B}_{D^{(*)}}\mathcal{B}_{K_S^0}$ is the product of the intermediate branching fractions of the relevant $D^{(*)}$ and $K_S^0$ decays in the reconstructed decay chain, $N_{B\overline B}$ is the total number of $B\overline B$ pairs, and $f_{+-,00}$ is the fraction of charged or neutral $B\overline B$ pairs.

The $f_{+-,00}$ values used in Eq.~\ref{eq:BR} are evaluated from the ratio ${f_{+-}/f_{00}}$ from Ref.~\cite{Belle:fpm00}, while ${N_{B\overline B}}$ is reported in Sec.~\ref{sec:Belle2}. 
We obtain ${2f_{\pm} N_{B\overline B}=(399\pm 11)\times 10^6}$ and ${2f_{0} N_{B\overline B}=(375\pm 10)\times 10^6}$. The branching fractions of the intermediate $D^{(*)}$ and $K_S^0$ decays are taken from the world-average values in Ref.~\cite{PDG}.

The signal yield as a function of $m(K^-K^{0}_S)$ is extracted using an $s$Plot, since $\Delta E$ and $m(K^-K^{0}_S)$ are uncorrelated. The $\Delta E$ distribution, fitted as described in Sec.~\ref{sec:yield}, is used as a discriminating variable to obtain the $s$Weights. The $s$Weights for the signal are then used to obtain the  $m(K^-K^{0}_S)$ distribution for signal only.  The efficiency $\varepsilon(m_{KK})$ described in Sec.~\ref{sec:efficiency} is used to extract the branching fraction. The results are given in Table~\ref{tab:BR_data}.

\begin{figure}[!hbt]
\centering
\subfigure{\includegraphics[width=0.49\columnwidth]{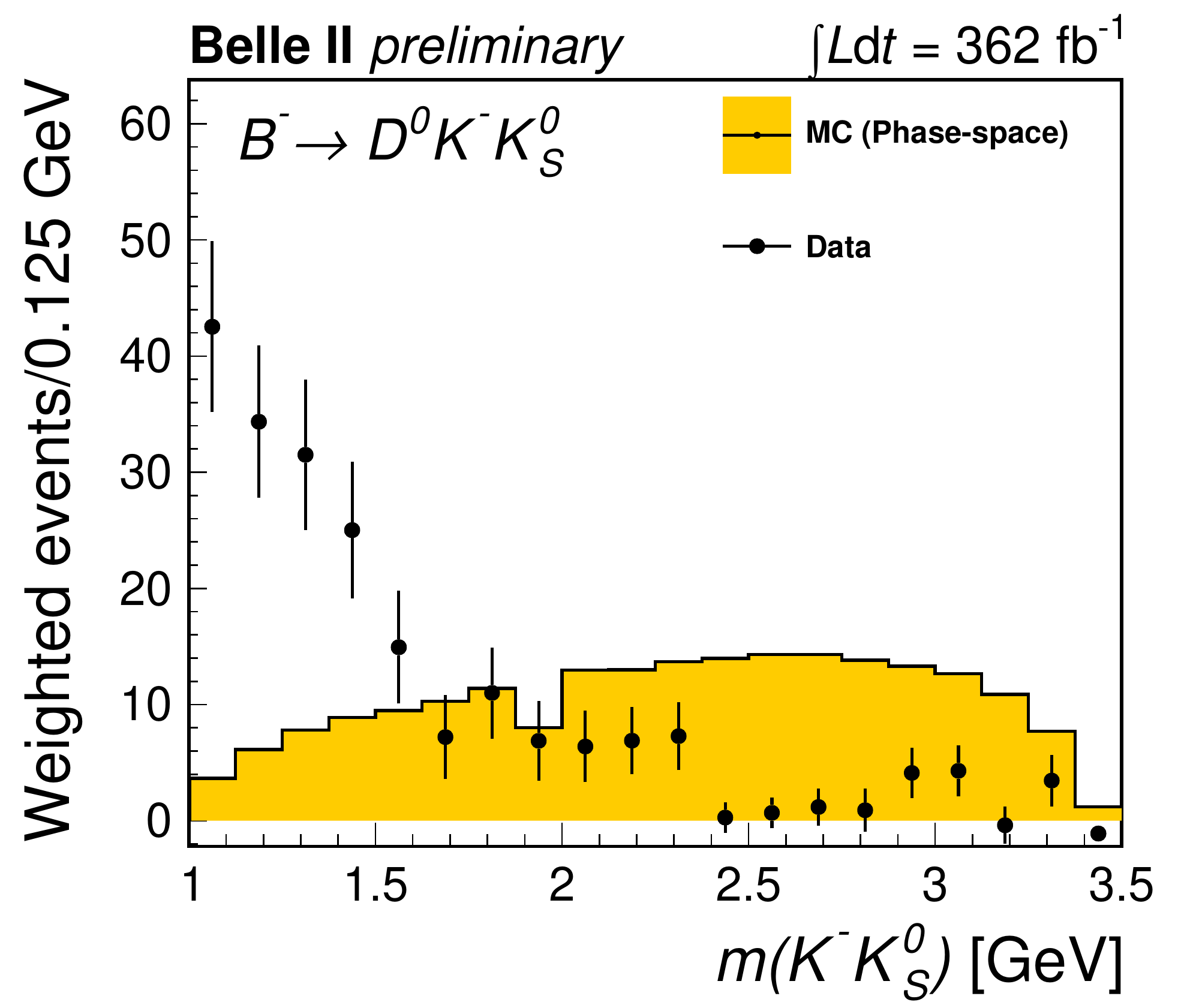}}
\subfigure{\includegraphics[width=0.49\columnwidth]{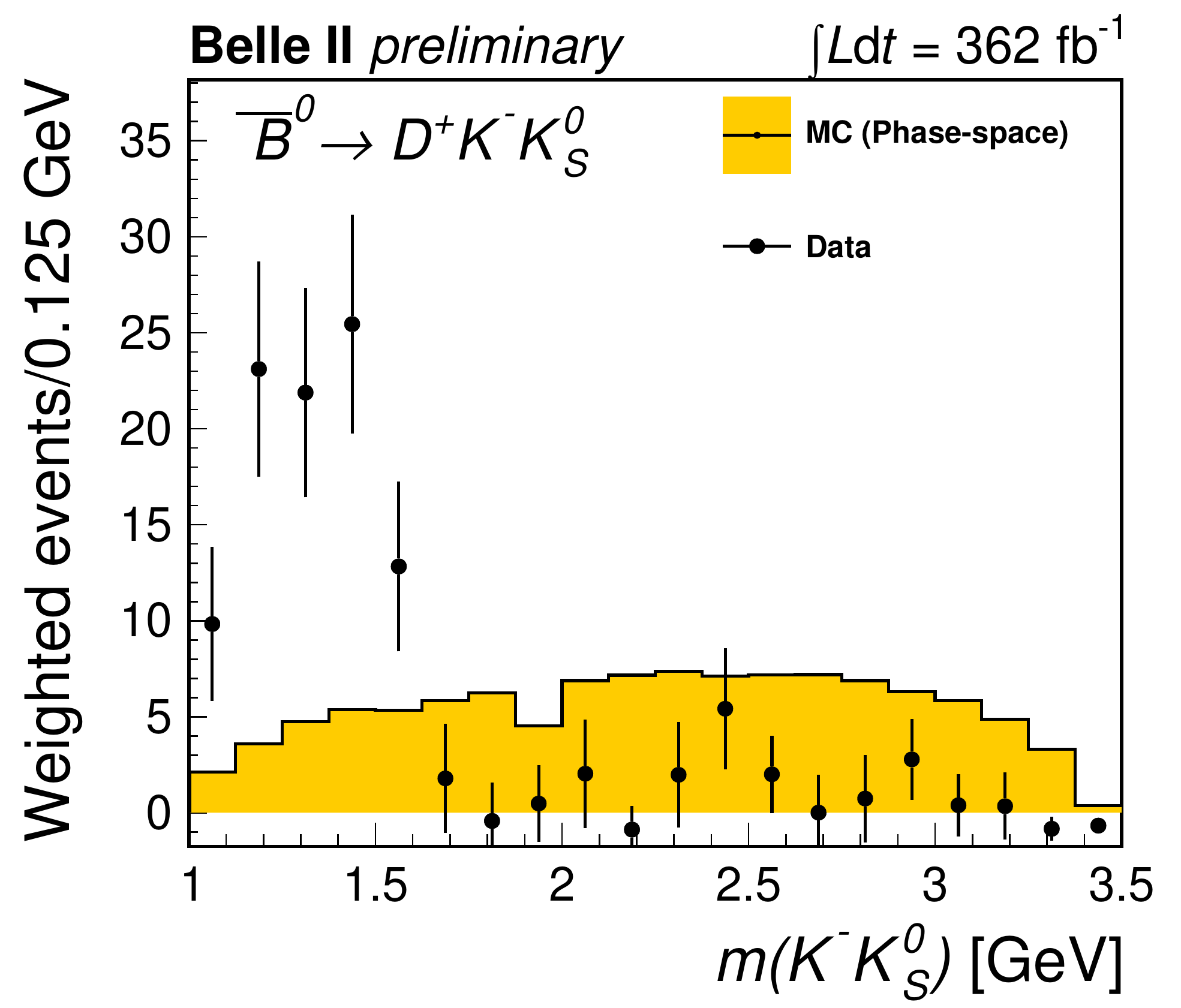}}
\subfigure{\includegraphics[width=0.49\columnwidth]{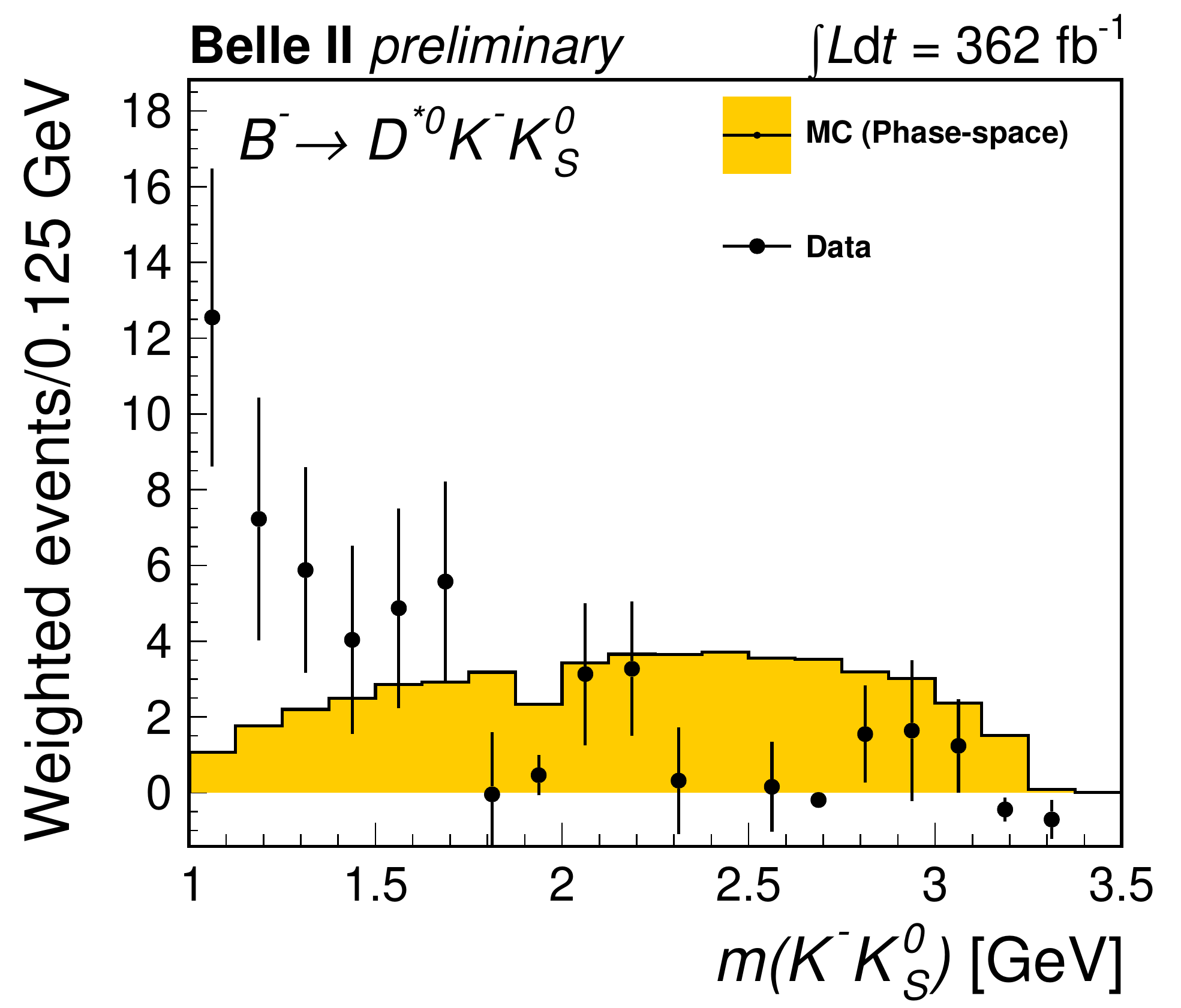}}
\subfigure{\includegraphics[width=0.49\columnwidth]{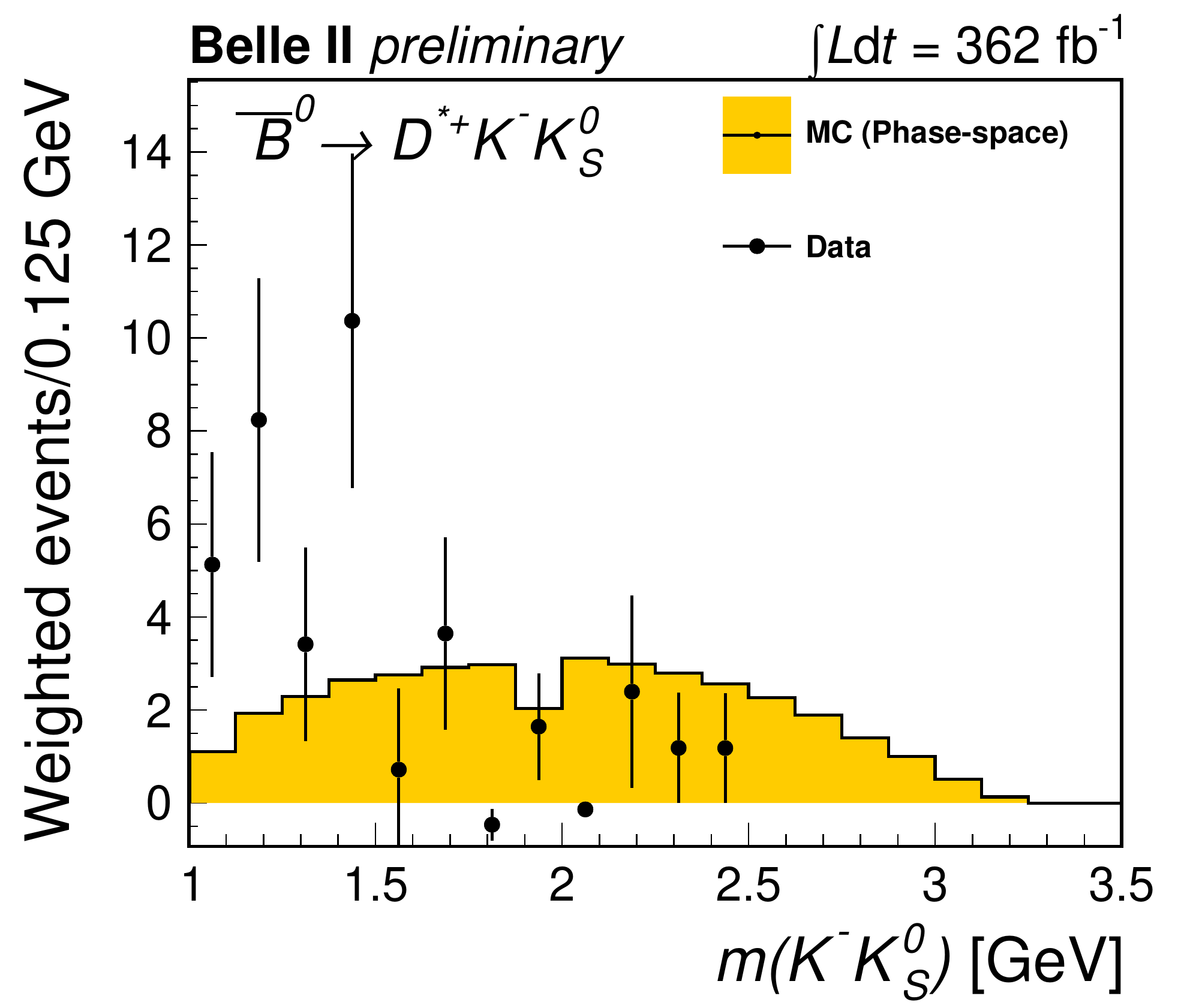}}
\caption{Distribution of $m(K^-K^{0}_S)$ for $B^-\to D^0K^-K_S^0$ (top left), $\overline B{}^0\to D^+K^-K_S^0$ (top right), $B^-\to D^{*0}K^-K_S^0$ (bottom left), and $\overline B{}^0\to D^{*+}K^-K_S^0$ (bottom right) channels. The background is subtracted by applying the signal $s$Weights. A phase-space MC simulation, rescaled to the integral of the data distribution, is also shown for comparison.} \label{fig:mKK_splot_KS}
\end{figure}

\begin{figure}[!hbt]
\centering
\subfigure{\includegraphics[width=0.24\columnwidth]{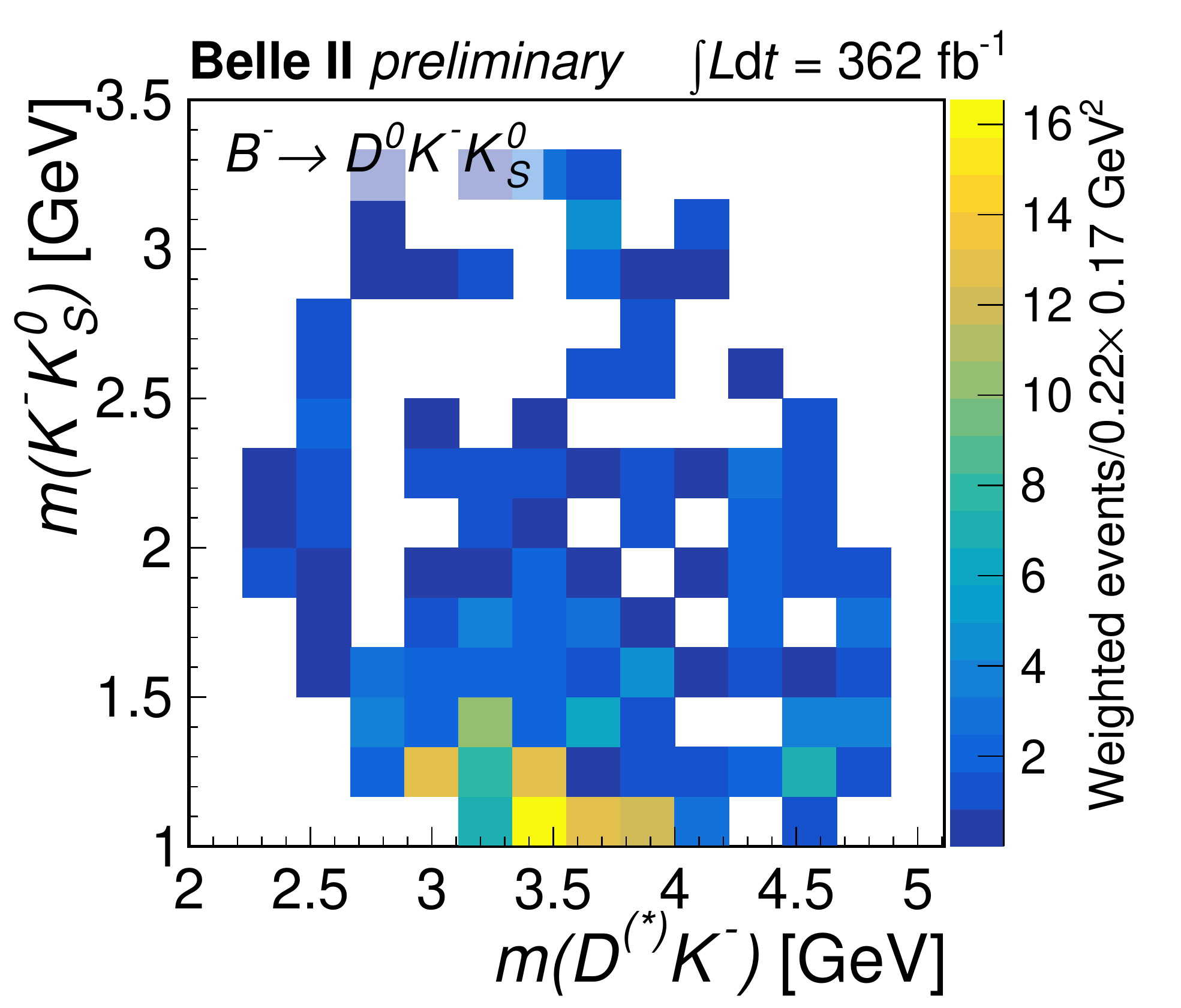}}
\subfigure{\includegraphics[width=0.24\columnwidth]{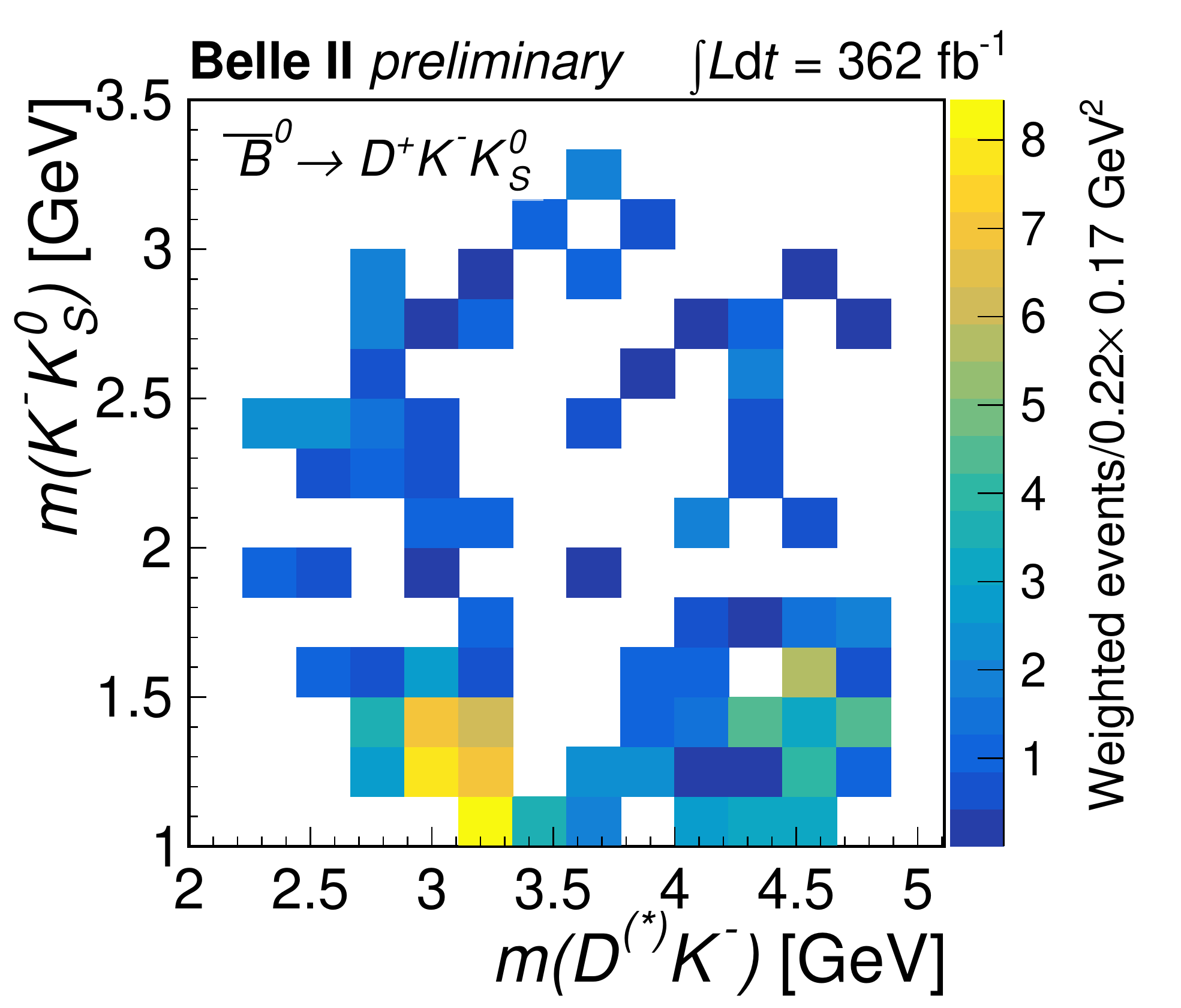}}
\subfigure{\includegraphics[width=0.24\columnwidth]{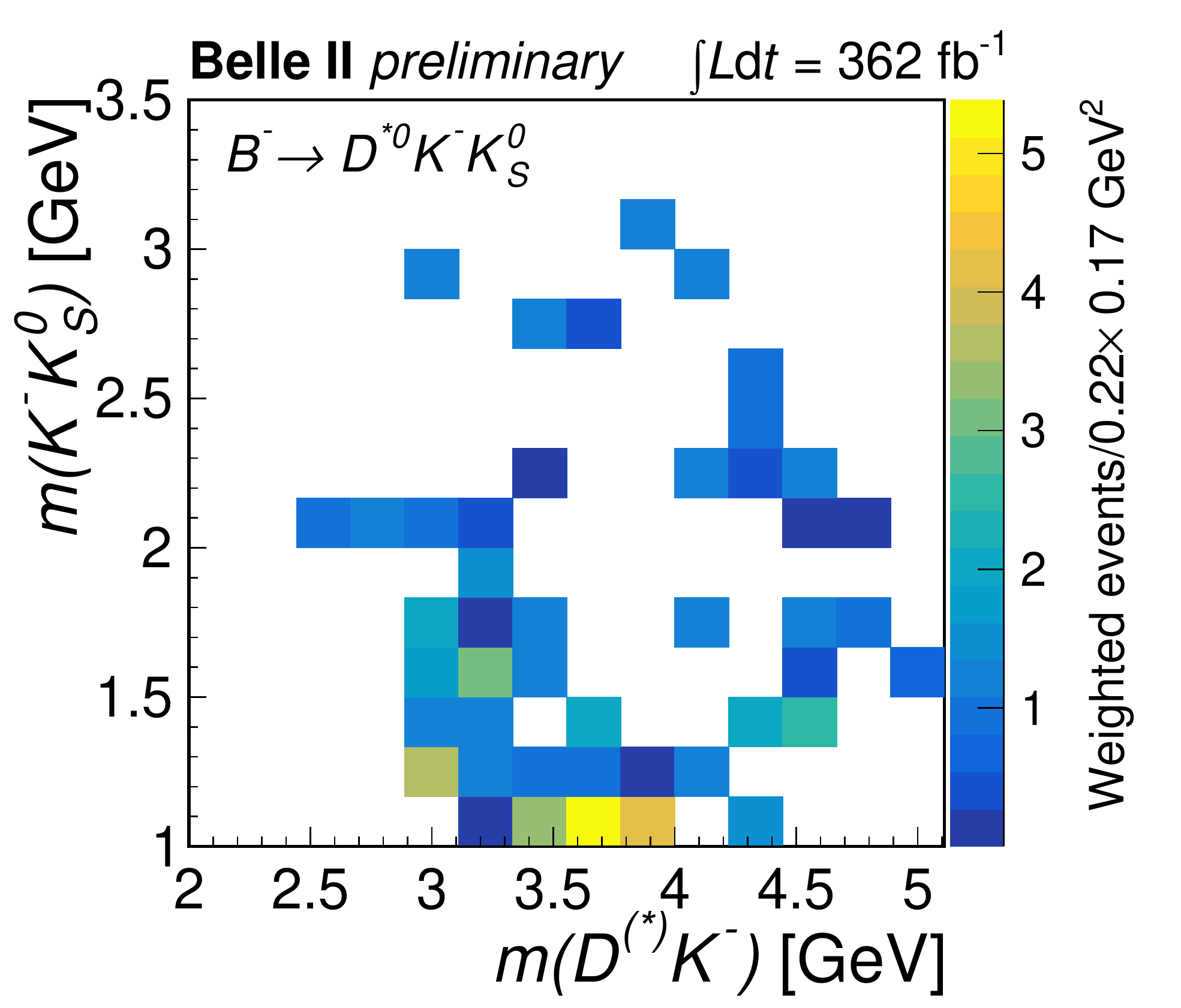}}
\subfigure{\includegraphics[width=0.24\columnwidth]{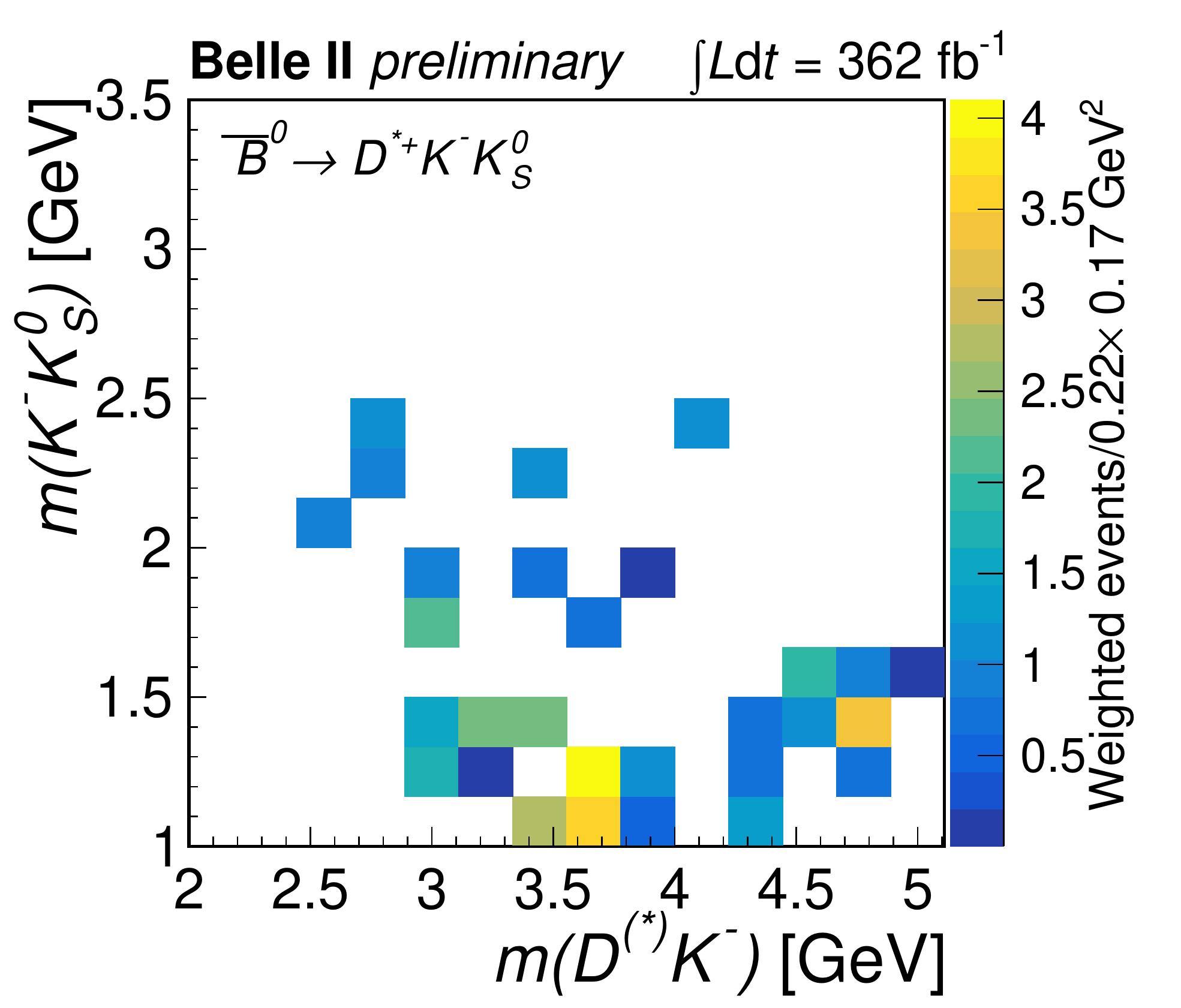}}
\subfigure{\includegraphics[width=0.24\columnwidth]{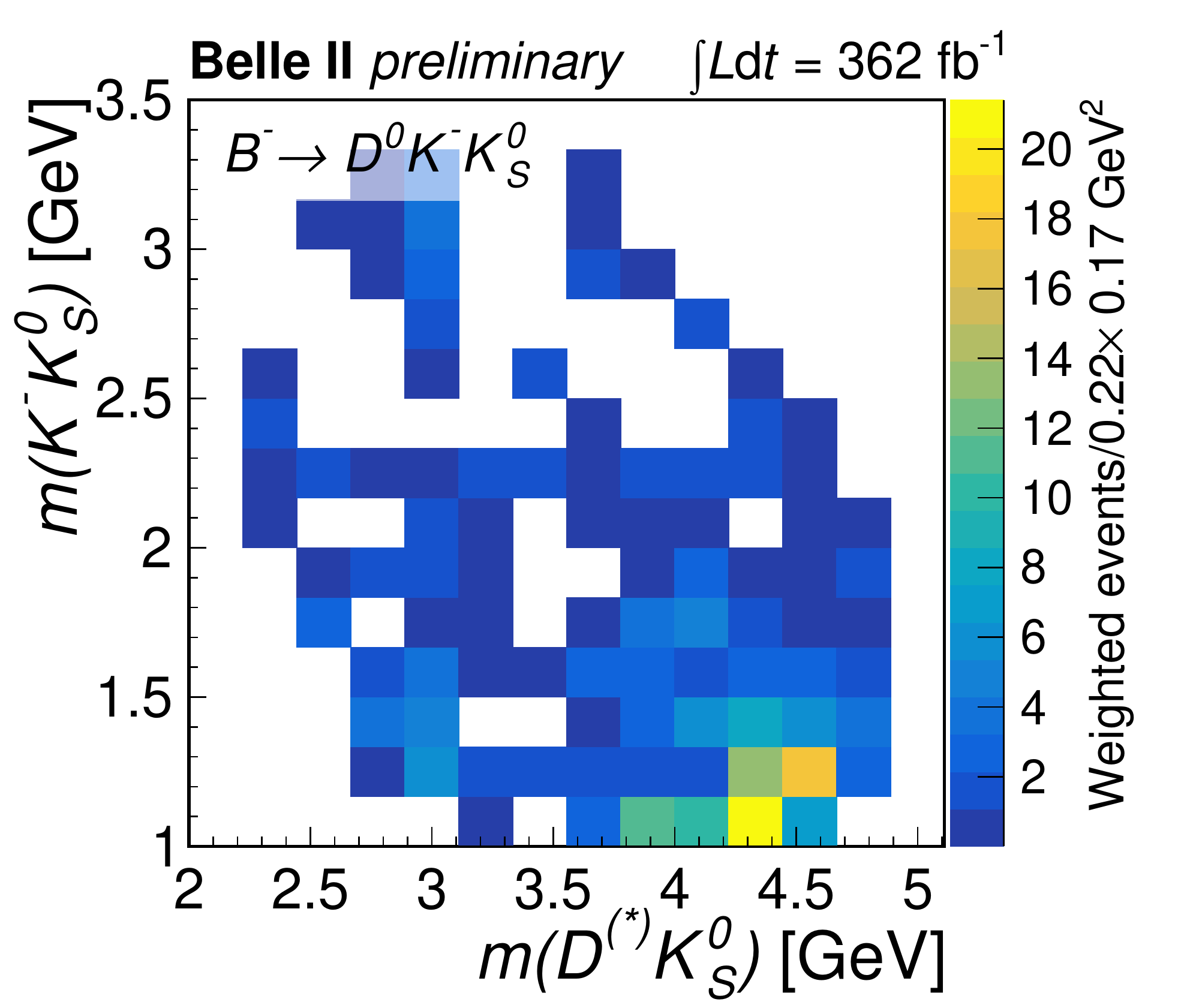}}
\subfigure{\includegraphics[width=0.24\columnwidth]{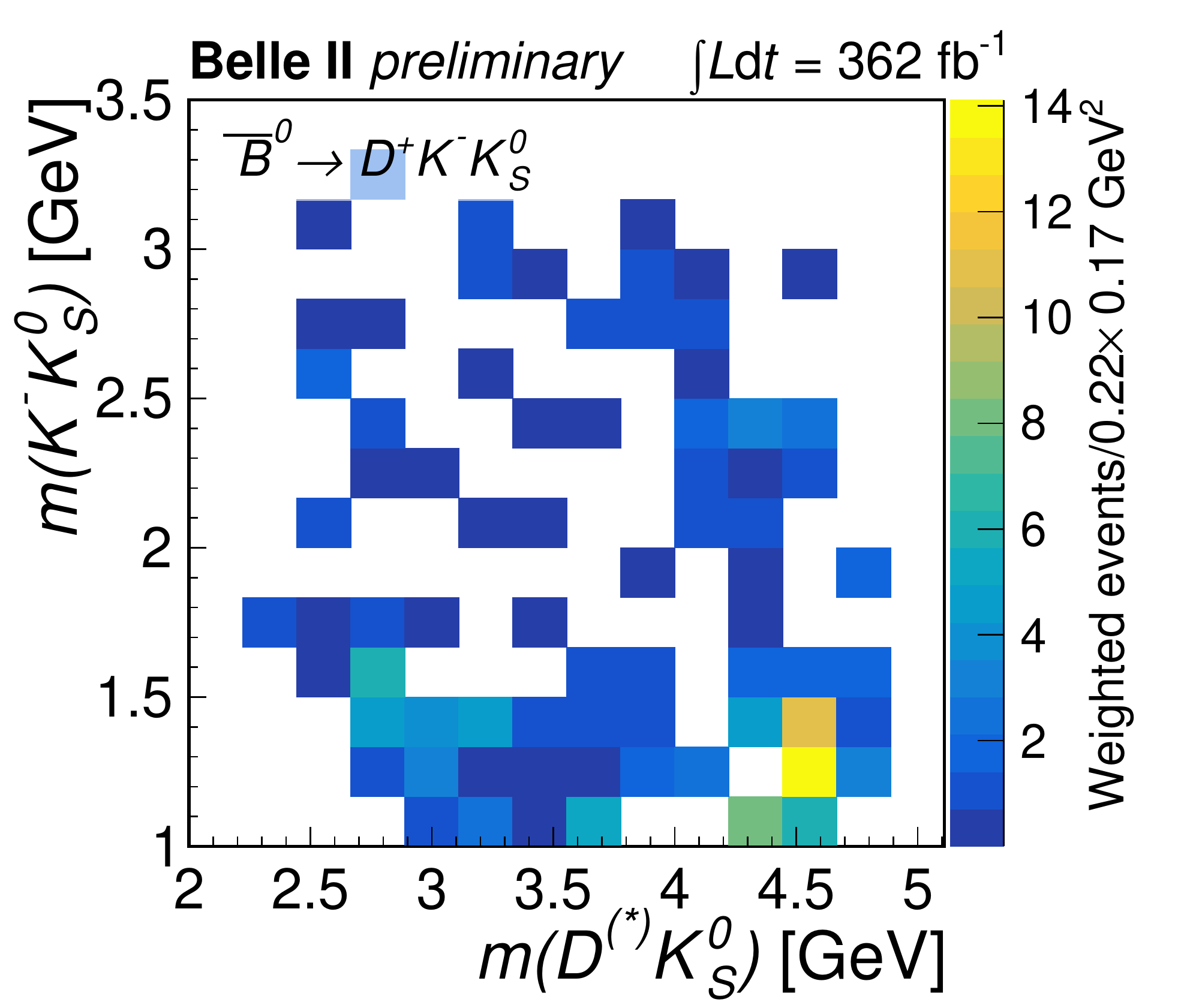}}
\subfigure{\includegraphics[width=0.24\columnwidth]{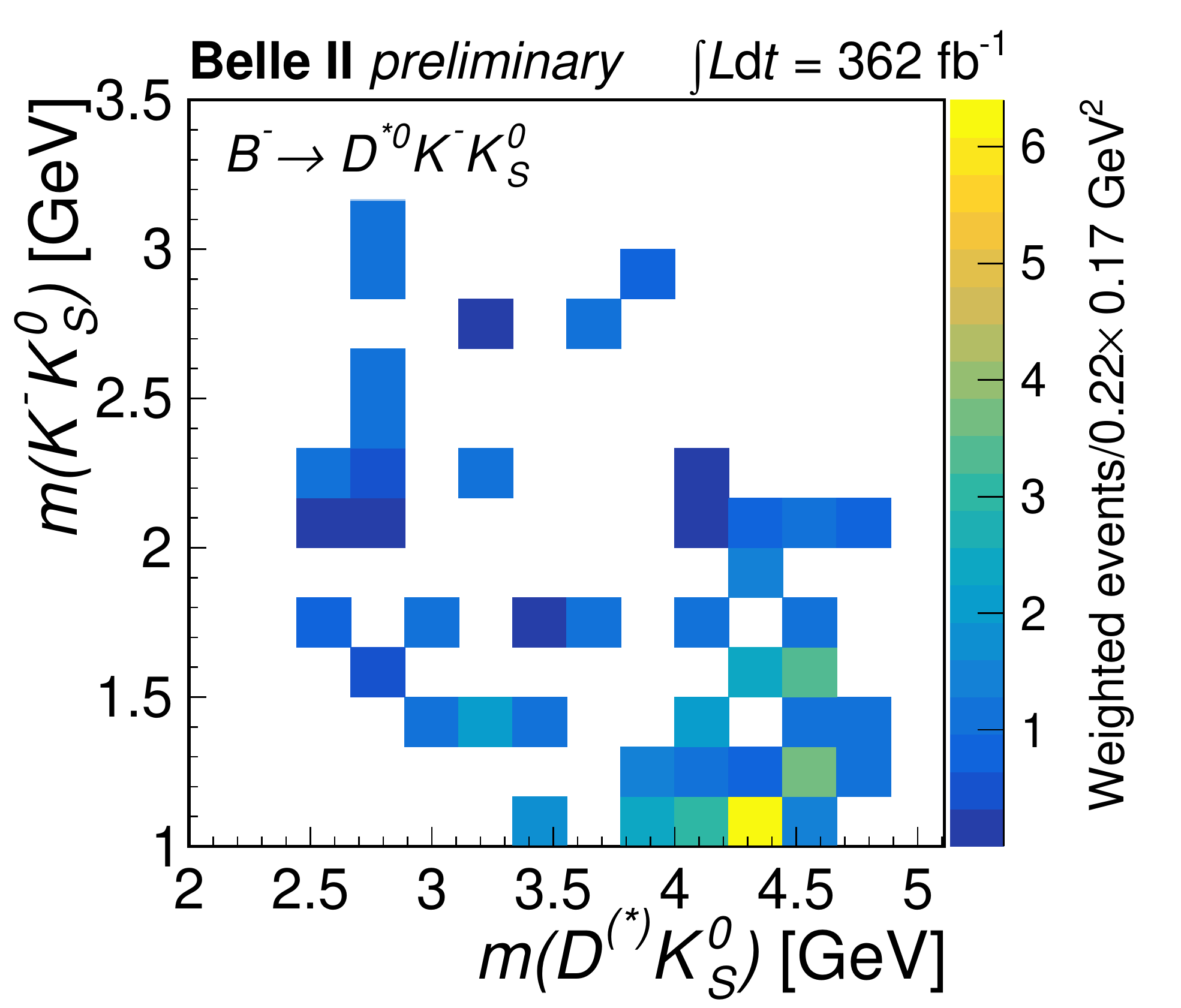}}
\subfigure{\includegraphics[width=0.24\columnwidth]{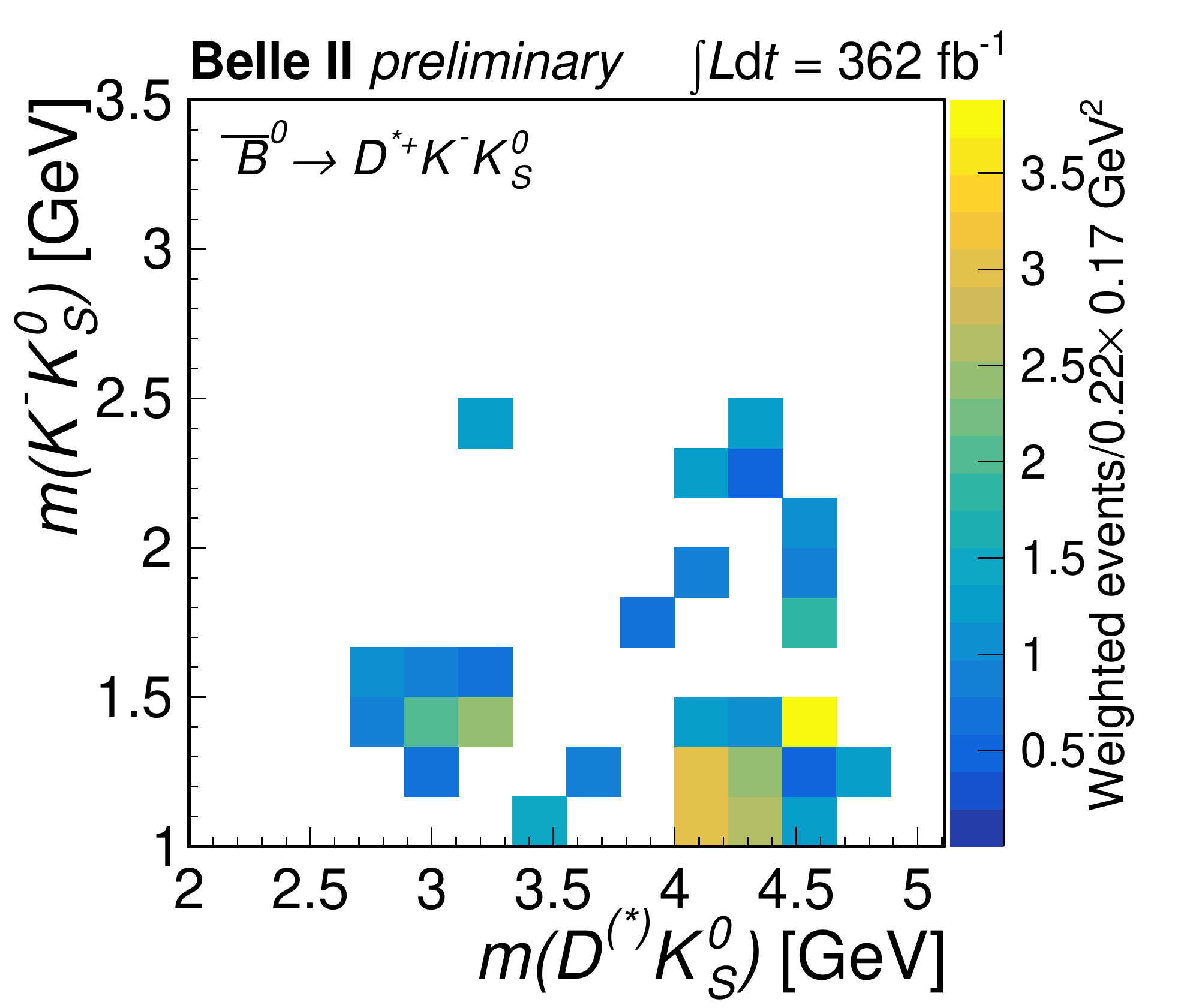}}
\caption{Dalitz distributions of $\bigl(m(D^{*}K^-), m(K^-K^{0}_S)\bigl)$ (upper panels) and $\bigl(m(D^{*}K_S^0), m(K^-K^{0}_S)\bigl)$ (lower panels)  for (left to right) $B^-\to D^0K^-K_S^0$, $\overline B{}^0\to D^+K^-K_S^0$, $B^-\to D^{*0}K^-K_S^0$, and $\overline B{}^0\to D^{*+}K^-K_S^0$ channels. The background is subtracted by applying the signal $s$Weight.} \label{fig:Dalitz}
\end{figure}

The data $m(K^-K^{0}_S)$ distributions are shown in Fig.~\ref{fig:mKK_splot_KS}. 
The $m(K^-K^{0}_S)$ distribution from the signal simulation is also overlaid to show the comparison with a phase-space distribution. In all channels, the bulk of the observed  $m(K^- K_S^0)$ distributions is located at low $m(K^- K_S^0)$, showing structures in all channels. The distributions vanish above 2.0--2.5~GeV, disfavouring the presence of a significant phase-space component. 
The $\bigl(m(D^{*}K^-), m(K^-K^{0}_S)\bigl)$ and $\bigl(m(D^{*}K_S^0), m(K^-K^{0}_S)\bigl)$ Dalitz distributions are shown in Fig.~\ref{fig:Dalitz} for the four channels. The selection is applied together with $s$Weights to subtract the background component. The Dalitz distributions also show some peaking structures that are not described by the simulation and require further investigation.

\begin{table}[!htb]
\caption{Analysis results. Measured yield for the four channels; average efficiency defined as ${\varepsilon=N_\text{reco}/\sum_{i\in m(K^-K^{0}_S)~\text{bins}}(N_i^\text{reco}/\varepsilon_i) }$ with the correction described in Sec.~\ref{sec:efficiency}; measured branching fractions (the first uncertainty is statistical, the second systematic).}\label{tab:BR_data}
\begin{tabularx}{1.0\linewidth}{XXXX}
\toprule
 Channel                               & Yield         & Average $\varepsilon$ & $\mathcal{B}$ [$10^{-4}$]        \\ 
\midrule
$B^- \to D^0 K^- K^0_S$               & $207 \pm 17 $  & $0.1009 \pm 0.0006$                 & $1.89 \pm 0.16 \pm 0.10$             \\
$\overline B{}^0 \to D^{+} K^- K^{0}_S$ & $108 \pm 14$ & $0.0525 \pm 0.0005$                & $0.85 \pm 0.11 \pm 0.05$               \\
$B^- \to D^{*0} K^- K^0_S$            & $51 \pm 9$  & $0.0457 \pm 0.0004$                & $1.57 \pm 0.27 \pm 0.12$                \\
$\overline B{}^0 \to D^{*+} K^- K^0_S$  & $37 \pm 7$  & $0.0563 \pm 0.0004$                & $0.96 \pm 0.18 \pm 0.06$               \\ 
\bottomrule
\end{tabularx}
\end{table}

\pagebreak
\section{Systematic Uncertainties}\label{sec:syst}

The contributions to the statistical and systematic uncertainties for each channel are summarized in Table~\ref{tab:systematicsRel}, expressed as relative uncertainties.  

The first group of systematic uncertainties affects the efficiency estimation. They are related to the disagreement between efficiencies in data and in simulation, and they are derived from the uncertainties on the corresponding corrections. 
An uncertainty associated with the limited size of the MC-simulation sample is included, which gives a statistical uncertainty in the efficiency estimation. 
The uncertainty related to the tracking efficiency is estimated using a $e^+e^-\to \tau^+\tau^-$ control sample.  No correction is applied, but a per-track uncertainty of 0.24\% is included assuming full correlation between the tracks. 
For the uncertainty associated with the  ${K_S^0}$ efficiency, a scale factor determined in $D^{*+}\to \overline D{}^0(\to K_S^0\pi^+\pi^-) \pi^+$ decays is provided with a fully correlated uncertainty equal to $\pm 100\%$ of the correction ($0.55\%/\text{cm}$ times the mean displacement of the decay vertex). 
For the uncertainty associated with the PID-efficiency correction, two variations of the nominal scale factors, according to the uncertainty, are applied to the efficiency. This uncertainty is assumed to be fully correlated between tracks and in the $p\times \theta$ plane of the tracks. The scale factors and the associated uncertainty are evaluated using $D^{*+}\to \overline D{}^0(\to K^-\pi^+) \pi^+$ and $K_S^0\to \pi^+\pi^-$ control samples. 
For the uncertainty associated with the low-momentum charged pion efficiency, the scale factors are provided with statistical (partially correlated in momentum) and systematic (correlated) uncertainties evaluated on a $B^0\to D^{*-}\pi^+$ control sample. A fully correlated 0.027 uncertainty is used for the scale factors, equal to the sum in quadrature of both uncertainties. This uncertainty affects only the $D^{*+}$ channel. 
For the uncertainty associated with the ${\pi^0}$-efficiency correction, a variation of the scale factor is defined according to the uncertainties evaluated on a $B^+\to D^{*0}\pi^+$ control sample. This uncertainty affects only the $D^{*0}$ channels. 

The fit model of the signal is modified to check its robustness.  Alternative branching fractions are evaluated, and the systematic uncertainty is quoted as the difference between the nominal and the alternative branching fraction. The fit model developed on signal simulation contains a constant function in addition to the core and tail Gaussians, to avoid being biased by outliers. An alternative model of the fit is tested without the constant function, but restricting to the range $-0.1~\text{GeV}<\Delta E<0.1~\text{GeV}$. 
A second alternative model is also tested, replacing the tail Gaussian with a Crystal-Ball function~\cite{CrystalBall}, with mean common to that of the core Gaussian function and the remaining shape parameters fixed from the signal simulation fit. 
A third variation is considered for the channels in which the fudge factors are fixed using auxiliary channels. The systematic uncertainty is evaluated by shifting the fudge factor within its uncertainty ($\delta F = 0.1$ for $D^{(*)+}$ and $\delta F = 0.03$ for $D^{*0}$) and repeating the fit. This variation is applied to the $D^{+}$, $D^{*0}$, and $D^{*+}$ channels. 
A fourth variation is obtained with a fit to the $B\to D^{(*)}D_s^-$ sample performed on data allowing the widths of the tail Gaussian free in the fit. The difference between the fixed values (i.e. those fixed from signal simulation samples) and the values obtained from the fit are used as variations of the fixed values of the widths in the $B\to D^{(*)}K^-K_S^0$ fit, to obtain an alternative value of branching fraction. 
Given their small signal yields, the latter approach is not feasible for the $\overline B{}^0\to D^+D_s^-$ and $B\to D^{*}D_s^-$ modes. Therefore, the variation observed in the $B^-\to D^0D_s^-$  mode is included for the $D^+$ and $D^{*+}$ modes, while for the $D^{*0}$ mode this variation is not included, but a different one is added as described below.

The $B^-\to D^{*0}K^-K_S^0$ signal includes a 29\% self-cross-feed component (i.e. misreconstructed signal events, mostly due to incorrect $\pi^0$ associations). Since the self-cross-feed component is composed of signal events, it does not artificially increase the branching ratio of the signal channels. However, it does degrade the resolution in $\Delta E$. Given the possible disagreement in the description of the $\pi^0$ misreconstruction between data and simulation, a dedicated systematic uncertainty is assigned. 
A fit to the $B^-\to D^{*0}\pi^-$ control channel is performed on data allowing the widths of the tail Gaussian $\sigma_L$ and $\sigma_R$ free in the fit, and on the simulation fixing the values of the widths. The difference in the widths resulting from two fits is used as variation to obtain an alternative value of $\mathcal B$. 
The systematic uncertainty is the difference between the nominal and the varied branching fraction. The self-cross-feed is small in the other three channels, thus the self-cross-feed systematic uncertainty is included in the fit-model systematic uncertainty.
    
A systematic uncertainty related to the specific choice of background model is assigned. In the nominal fit, the background, for all channels, is described with the sum of an exponential function and a constant. Two alternative fits are performed by adding a linear term or removing the constant term. The differences between the nominal branching fractions and the results obtained with the alternative background models are quoted as systematic uncertainties.
    
A systematic uncertainty is assigned to account for the uncertainties in Eq.~\ref{eq:DstpBkg_yield}, used to assess the yield of the feed-across from ${\overline B{}^0\to D^{*+}K^-K^0_S}$ channel in the $D^{*0}$ channel. Equation~\ref{eq:Dst0_hypothesis} is verified with signal simulation, and holds with a maximum deviation of 10\%. The yield of the peaking background $N_{D^{*+}}$ is scaled to 110\% and 90\% and two resulting signal yields are extracted. 
In addition, the branching fractions of $D^{*+}$ and $D^{0}$ channels are scaled by their uncertainties, to take into account the correlation between the branching fractions and the $N_{D^{*+}}$ yield, producing additional variations of $\mathcal B(\overline{}B^{0}\to D^{*0}K^-K_S^0)$. The systematic uncertainty is the sum in quadrature of the differences between the alternative branching fractions and the nominal one. This systematic uncertainty affects only the $D^{*0}$ channels. 

A systematic uncertainty related to the $\delta N_{B\overline B}=5.6\times 10^6$ uncertainty in the total number of $B\overline B$ pairs, which enters in Eq.~\ref{eq:BR},  is considered and combined with the uncertainty on $f_{+-,00}$~\cite{Belle:fpm00}. 

A systematic uncertainty related to the intermediate branching fraction used in Eq.~\ref{eq:BR} is also considered propagating the uncertainties of the known intermediate branching fractions to the measured branching fraction.

\begin{table}[!hbt]
\caption{Relative statistical uncertainties and breakdown of the relative systematic uncertainties for the four channels. All the values are in percent. A dash is placed where the uncertainty is not applicable. }\label{tab:systematicsRel}
\hspace{-1cm}
\begin{tabularx}{1.1\linewidth}{l c c c c }

\toprule
Source  & $B^- \to D^0 K^- K^0_S$ & $\overline B{}^0 \to D^{+} K^- K^{0}_S$& $B^- \to D^{*0} K^- K^0_S$  & $\overline B{}^0 \to D^{*+} K^- K^0_S$  \\
\midrule
Eff. - MC sample size  & 0.6 & 0.9 & 1.0 & 0.8 \\

Eff. -  tracking & 0.7 & 1.0 & 0.7 & 1.0 \\

Eff. - $\pi^+$ from $D^{*+}$  & - & - & - & 2.7 \\ 

Eff. - $K_S^0$  & 3.4 & 3.4 & 3.4 & 3.3\\

Eff. -  PID  & 1.3 & 1.4 & 0.5 & 0.6 \\

Eff. -  $\pi^0$ & - & - &  5.1 & - \\

 Signal model & 1.9 & 3.3 & 2.7 & 3.1\\

 Bkg model & 1.1 & 0.8 & 0.1 & 0.1 \\

 Self-cross-feed & - & - & 2.7 & - \\
 
 $D^{*0}$ peaking bkg & - & - & 0.9 & - \\
 
$N_{B\overline B}$, $f_{+-,00}$  & 2.7 & 2.8 & 2.7 & 2.8 \\
 
Intermediate $\mathcal{B}$s  & 0.7 & 1.7 & 1.6 & 1.1\\
\midrule
Total systematic  & 5.2 & 6.1 & 7.6 & 6.2\\
 \midrule
 Statistical        & 8.3 & 13.5 & 17.1 & 19.0 \\
\bottomrule
\end{tabularx}
\end{table}

\section{Conclusions}

We report the first observation of three new $B$ decay channels: $\overline B{}^0\to D^+K^-K_S^0$,  $B^-\to D^{*0}K^-K_S^0$, and $\overline B{}^0\to D^{*+}K^-K_S^0$. The branching fractions of the three aforementioned decays and of $B^-\to D^0K^-K_S^0$ are found to be
\begin{align*}
    \mathcal{B}(B^-\to D^0K^-K_S^0)=&(1.89\pm 0.16\pm 0.10)\times 10^{-4},\\
    \mathcal{B}(\overline B{}^0\to D^+K^-K_S^0)=&(0.85\pm 0.11\pm 0.05)\times 10^{-4},\\
    \mathcal{B}(B^-\to D^{*0}K^-K_S^0)=&(1.57\pm 0.27\pm 0.12)\times 10^{-4},\\
    \mathcal{B}(\overline B{}^0\to D^{*+}K^-K_S^0)=&(0.96\pm 0.18\pm 0.06)\times 10^{-4},
\end{align*} 
where the first uncertainty is statistical and the second systematic. 
The results are based on an electron-positron data sample collected by Belle II on the $\Upsilon(4S)$ resonance with an integrated luminosity of $362~fb^{-1}$. 
The precision of the four measurements is between 10\% and 20\% and is limited by statistical uncertainties. 
The precision of $\mathcal{B}(B^-\to D^0K^-K_S^0)$ is improved by more than a factor of three compared to previous measurements~\cite{Belle:DKK}. 

The observed $m(K^-K^{0}_S)$ distribution for the four channels is far from a three-body phase-space distribution. It shows a peaking structure in the low-invariant mass region in all channels. This structure may indicate the presence of a resonant component, as has been previously reported by Belle~\cite{Belle:DKK}. The three-body Dalitz distributions also show some structures. Further investigation is needed to explain the observed distributions, and to measure the resonant and non-resonant branching fractions. 

\FloatBarrier

\section*{Acknowledgements}
This work, based on data collected using the Belle II detector, which was built and commissioned prior to March 2019, was supported by
Science Committee of the Republic of Armenia Grant No.~20TTCG-1C010;
Australian Research Council and research Grants
No.~DE220100462,
No.~DP180102629,
No.~DP170102389,
No.~DP170102204,
No.~DP150103061,
No.~FT130100303,
No.~FT130100018,
and
No.~FT120100745;
Austrian Federal Ministry of Education, Science and Research,
Austrian Science Fund
No.~P~31361-N36
and
No.~J4625-N,
and
Horizon 2020 ERC Starting Grant No.~947006 ``InterLeptons'';
Natural Sciences and Engineering Research Council of Canada, Compute Canada and CANARIE;
Chinese Academy of Sciences and research Grant No.~QYZDJ-SSW-SLH011,
National Natural Science Foundation of China and research Grants
No.~11521505,
No.~11575017,
No.~11675166,
No.~11761141009,
No.~11705209,
and
No.~11975076,
LiaoNing Revitalization Talents Program under Contract No.~XLYC1807135,
Shanghai Pujiang Program under Grant No.~18PJ1401000,
Shandong Provincial Natural Science Foundation Project~ZR2022JQ02,
and the CAS Center for Excellence in Particle Physics (CCEPP);
the Ministry of Education, Youth, and Sports of the Czech Republic under Contract No.~LTT17020 and
Charles University Grant No.~SVV 260448 and
the Czech Science Foundation Grant No.~22-18469S;
European Research Council, Seventh Framework PIEF-GA-2013-622527,
Horizon 2020 ERC-Advanced Grants No.~267104 and No.~884719,
Horizon 2020 ERC-Consolidator Grant No.~819127,
Horizon 2020 Marie Sklodowska-Curie Grant Agreement No.~700525 "NIOBE"
and
No.~101026516,
and
Horizon 2020 Marie Sklodowska-Curie RISE project JENNIFER2 Grant Agreement No.~822070 (European grants);
L'Institut National de Physique Nucl\'{e}aire et de Physique des Particules (IN2P3) du CNRS (France);
BMBF, DFG, HGF, MPG, and AvH Foundation (Germany);
Department of Atomic Energy under Project Identification No.~RTI 4002 and Department of Science and Technology (India);
Israel Science Foundation Grant No.~2476/17,
U.S.-Israel Binational Science Foundation Grant No.~2016113, and
Israel Ministry of Science Grant No.~3-16543;
Istituto Nazionale di Fisica Nucleare and the research grants BELLE2;
Japan Society for the Promotion of Science, Grant-in-Aid for Scientific Research Grants
No.~16H03968,
No.~16H03993,
No.~16H06492,
No.~16K05323,
No.~17H01133,
No.~17H05405,
No.~18K03621,
No.~18H03710,
No.~18H05226,
No.~19H00682, 
No.~22H00144,
No.~26220706,
and
No.~26400255,
the National Institute of Informatics, and Science Information NETwork 5 (SINET5), 
and
the Ministry of Education, Culture, Sports, Science, and Technology (MEXT) of Japan;  
National Research Foundation (NRF) of Korea Grants
No.~2016R1\-D1A1B\-02012900,
No.~2018R1\-A2B\-3003643,
No.~2018R1\-A6A1A\-06024970,
No.~2018R1\-D1A1B\-07047294,
No.~2019R1\-I1A3A\-01058933,
No.~2022R1\-A2C\-1003993,
and
No.~RS-2022-00197659,
Radiation Science Research Institute,
Foreign Large-size Research Facility Application Supporting project,
the Global Science Experimental Data Hub Center of the Korea Institute of Science and Technology Information
and
KREONET/GLORIAD;
Universiti Malaya RU grant, Akademi Sains Malaysia, and Ministry of Education Malaysia;
Frontiers of Science Program Contracts
No.~FOINS-296,
No.~CB-221329,
No.~CB-236394,
No.~CB-254409,
and
No.~CB-180023, and No.~SEP-CINVESTAV research Grant No.~237 (Mexico);
the Polish Ministry of Science and Higher Education and the National Science Center;
the Ministry of Science and Higher Education of the Russian Federation,
Agreement No.~14.W03.31.0026, and
the HSE University Basic Research Program, Moscow;
University of Tabuk research Grants
No.~S-0256-1438 and No.~S-0280-1439 (Saudi Arabia);
Slovenian Research Agency and research Grants
No.~J1-9124
and
No.~P1-0135;
Agencia Estatal de Investigacion, Spain
Grant No.~RYC2020-029875-I
and
Generalitat Valenciana, Spain
Grant No.~CIDEGENT/2018/020
Ministry of Science and Technology and research Grants
No.~MOST106-2112-M-002-005-MY3
and
No.~MOST107-2119-M-002-035-MY3,
and the Ministry of Education (Taiwan);
Thailand Center of Excellence in Physics;
TUBITAK ULAKBIM (Turkey);
National Research Foundation of Ukraine, project No.~2020.02/0257,
and
Ministry of Education and Science of Ukraine;
the U.S. National Science Foundation and research Grants
No.~PHY-1913789 
and
No.~PHY-2111604, 
and the U.S. Department of Energy and research Awards
No.~DE-AC06-76RLO1830, 
No.~DE-SC0007983, 
No.~DE-SC0009824, 
No.~DE-SC0009973, 
No.~DE-SC0010007, 
No.~DE-SC0010073, 
No.~DE-SC0010118, 
No.~DE-SC0010504, 
No.~DE-SC0011784, 
No.~DE-SC0012704, 
No.~DE-SC0019230, 
No.~DE-SC0021274, 
No.~DE-SC0022350; 
and
the Vietnam Academy of Science and Technology (VAST) under Grant No.~DL0000.05/21-23.

These acknowledgements are not to be interpreted as an endorsement of any statement made
by any of our institutes, funding agencies, governments, or their representatives.

We thank the SuperKEKB team for delivering high-luminosity collisions;
the KEK cryogenics group for the efficient operation of the detector solenoid magnet;
the KEK computer group and the NII for on-site computing support and SINET6 network support;
and the raw-data centers at BNL, DESY, GridKa, IN2P3, INFN, and the University of Victoria for offsite computing support.

\printbibliography[heading=bibintoc]

\clearpage

\end{document}